\documentclass[showkeys,superscriptaddress,floatfix,prd,10pt,aps]{revtex4-2}
\usepackage{graphicx,epstopdf}
\usepackage[caption=false]{subfig}
\usepackage{appendix}
\usepackage[T1]{fontenc}
\usepackage{lmodern}
\usepackage[dvipsnames,x11names]{xcolor}
\usepackage[colorlinks=true,linkcolor=ForestGreen,citecolor=ForestGreen,urlcolor=NavyBlue]{hyperref}
\usepackage[sort&compress]{natbib}
\usepackage{lipsum}
\usepackage{morefloats}
\usepackage[pdf]{pstricks}
\usepackage{amsmath}
\usepackage{amssymb}
\usepackage{amsfonts}
\usepackage{rotating}
\usepackage{cancel}
\usepackage{mathtools}
\usepackage{bbm}
\usepackage{dsfont}
\usepackage{bbold}
\usepackage{multirow}
\usepackage{ulem}
\usepackage{physics}
\usepackage{orcidlink}
\usepackage{colortbl}

\def\pslash{p\!\!\!\slash }
\def\qslash{q\!\!\!\slash }
\def\xslash{x\!\!\!\slash }
\def\eslash{\varepsilon\!\!\!\slash }

\def\vel{\left|}
\def\ver{\right|}

\begin{document}

\title{Electromagnetic properties of $\Omega_{c}^0$ resonances via light-cone QCD}

\author{Ula\c{s}~\"{O}zdem\orcidlink{0000-0002-1907-2894}}%
\email[]{ulasozdem@aydin.edu.tr }
\affiliation{ Health Services Vocational School of Higher Education, Istanbul Aydin University, Sefakoy-Kucukcekmece, 34295 Istanbul, T\"{u}rkiye}

 
\begin{abstract}
We systematically study the electromagnetic properties of controversial states whose internal structure is not elucidated and we try to offer a different point of view to unravel the internal structure of these states.  Inspired by the $\Omega_c$ states observed by the LHCb Collaboration, we study the electromagnetic properties of the $\Omega_c$ states as the compact diquark-diquark-antiquark pentaquarks with both $J^P = \frac{1}{2}^-$ and $J^P = \frac{3}{2}^-$ in the context of the QCD light-cone sum rule model. From the obtained numerical results, we conclude that the magnetic dipole moments of the $\Omega_c$ states can reflect their inner structures, which can be used to distinguish their spin-parity quantum numbers. Measuring the magnetic moment of the $\Omega_c$ states in future experimental facilities can be very helpful for understanding the internal organization and identifying the quantum numbers of these states.
\end{abstract}

\maketitle

\section{Motivation}

Many heavy baryon states have been discovered in recent years by experimental collaborations. One of the main challenges in non-perturbative QCD is to understand and shed light on the precise nature of these states.  Through further theoretical explorations involving the study of hadrons comprising a single heavy quark, it offers an exquisite basis to probe the dynamics of a light diquark in a heavy quark background, to enhance the understanding of the non-perturbative nature of QCD, and to test the predictions of different phenomenological models. In recent decades, there have been significant experimental advancements in the field of singly-charm/bottom baryons, resulting in a dramatic increase in the number of particles~\cite{ARGUS:1993vtm, CLEO:1994oxm, ARGUS:1997snv, E687:1993bax, CLEO:2000mbh, BaBar:2006itc, Belle:2006xni, LHCb:2017jym, Ammosov:1993pi, CLEO:1996czm, Belle:2004zjl, CLEO:1995amh, CLEO:1996zcj,E687:1998dwp, CLEO:2000ibb, CLEO:1999msf, LHCb:2020iby, BaBar:2007xtc, Belle:2006edu, BaBar:2007zjt, BaBar:2006pve, LHCb:2017uwr, LHCb:2012kxf, LHCb:2020lzx, LHCb:2019soc, LHCb:2018haf, CMS:2021rvl,  LHCb:2018vuc,  LHCb:2020xpu, LHCb:2020tqd}. As the data for the presence of some of these states is scarce and their internal structure, as well as quantum numbers, are not well defined, additional experimental exploration is therefore required. Therefore, researchers in hadron physics continue to study these topics through theoretical and experimental research, as they have not yet been fully understood.

In 2017, the LHCb collaboration studied the  $\Xi_c^+ K^-$  mass spectrum and observed five new narrow excited $\Omega_c$ states,
$\Omega_c(3000)$, $\Omega_c(3050)$, $\Omega_c(3066)$, $\Omega_c(3090)$, $\Omega_c(3119)$ \cite{LHCb:2017uwr}.  The parameters that have been measured are as follows
%
\begin{align}
  {\Omega_c(3000)}:  M &= 3000.4 \pm 0.2 \pm 0.1 \mbox{ MeV}\,, \,\,\,\,
  \Gamma = 4.5\pm0.6\pm0.3 \mbox{ MeV}\,, 
  \nonumber \\
  {\Omega_c(3050)}: M & = 3050.2 \pm 0.1 \pm 0.1 \mbox{ MeV},\,\,\,\,\,
   \Gamma = 0.8\pm0.2\pm0.1 \mbox{ MeV} \,, \,\, \nonumber \\
     {\Omega_c(3066)}: M &= 3065.6 \pm 0.1 \pm 0.3 \mbox{ MeV}\,, \,\,\,\, 
  \Gamma = 3.5\pm0.4\pm0.2 \mbox{ MeV} \,, \nonumber \\
    {\Omega_c(3090)}: M &= 3090.2 \pm 0.3 \pm 0.5 \mbox{ MeV}\,, \,\,\,\,
  \Gamma = 8.7\pm1.0\pm0.8 \mbox{ MeV}\,, \nonumber \\
  {\Omega_c(3119)}: M &= 3119.1 \pm 0.3 \pm 0.9 \mbox{ MeV}\,, \,\,\,\, 
  \Gamma = 1.1\pm0.8\pm0.4 \mbox{ MeV}\,.
\end{align}

Later, the former four states  $\Omega_c(3000)$, $\Omega_c(3050)$, $\Omega_c(3066)$ and $\Omega_c(3090)$ were confirmed by the  LHCb and Belle Collaborations~\cite{Belle:2017ext,LHCb:2023sxp}. The discovery of the LHCb collaboration has led to a new experimental situation, which requires a more detailed study of heavy baryons and their properties.   The discovery of these states, although their quantum numbers have not been determined, could provide new insights into QCD and its complex behavior. This could lead to a deeper understanding of the underlying properties of QCD. However, our information about their properties is still not sufficient and further suggestions for experimental exploration of $\Omega_c$ states should be addressed. The experimental discoveries were followed by various theoretical studies, investigating them in the conventional baryon, molecular, and compact pentaquark states to shed light on their exact nature and quantum numbers~(for details see the Refs.~\cite{Guo:2017jvc,Olsen:2017bmm,Brambilla:2019esw,Cheng:2021qpd,Chen:2022asf}). 

The literature review indicates that the majority of research has concentrated on computing the spectroscopic and decay parameters of these states. However, it is evident that relying exclusively on spectroscopic and decay parameters is insufficient for resolving the disputed nature of these states. Therefore, to determine the precise internal configurations of these states, additional studies such as obtaining electromagnetic multipole moments, radiative decays, and weak decays are required. 
To unveil the nature and internal structures of unconventional states, the physical quantities associated with the electromagnetic properties of hadrons are valuable parameters. The charge and magnetism distributions inside the hadrons can be studied using the electromagnetic form factors (FFs) and the resulting multipole moments. This information can be used to determine the distribution of quark-antiquark pairs (both valence and sea-quarks) and gluons within the hadron volume. It also provides valuable insights into the geometric forms, electric radii, and magnetic radii of the hadrons. With this motivation, in this work, we will derive the electromagnetic features of $\Omega_c$ states using the method of QCD light-cone sum rules~\cite{Chernyak:1990ag,Braun:1988qv,Balitsky:1989ry}. Throughout the analysis, these states are considered in the compact $[ss][dc] [\bar d]$, $[sd][sc][\bar d]$ and $[ds][sc] [\bar d]$ diquark-diquark-antiquark configurations, and are assumed to have both $J^P = \frac{1}{2}^-$ and $J^P = \frac{3}{2}^-$ quantum numbers.  There are several studies in the literature that have extracted the electromagnetic features of the singly-heavy nonconventional hadron states~\cite{Ozdem:2023edw,Ozdem:2023okg,Ozdem:2023eyz,Ozdem:2022ydv,Ozdem:2021vry,Azizi:2021aib,Azizi:2018jky,Azizi:2018mte}.

The article is structured as follows: we derive the QCD light-cone sum rules for the electromagnetic features of the $\Omega_c$ states in Sect. \ref{secII}. The numerical results of the magnetic dipole moments and discussions are presented in Sect. \ref{secIII}. Sect. \ref{secIV} is reserved for our summary and concluding remarks.

 \begin{widetext}
  
\section{Formalism}\label{secII}

To study the electromagnetic properties of hadrons in the low energy regime, reliable and efficient non-perturbative methods are needed. The QCD light-cone sum rules method is one of these methods and has been very successful in predicting the static and dynamical properties of both conventional and nonconventional hadrons. In this method, the correlation function, the key ingredient of the model, is calculated in terms of both hadronic parameters and QCD degrees of freedom.  The Borel transform and the continuum subtraction procedures are then applied to suppress the contribution of the effects caused by the continuum and higher states. Together with these procedures, the QCD light-cone sum rules for the physical parameter of interest are obtained via dispersion integrals and by using the quark-hadron duality assumption. 

 The study of the electromagnetic form factors of the $\Omega_c$ states ($\frac{1}{2}^- \rightarrow \Omega_c$, $\frac{3}{2}^- \rightarrow \Omega_c^*$) in QCD light cone sum rules begins with the writing of the correlation function in the weak electromagnetic background field ($\gamma$) as follows
 \begin{align} \label{edmn01}
 \Pi(p,q)&=i\int d^4x e^{ip \cdot x} \langle0|T\left\{J^{\Omega_c}(x)\bar{J}^{\Omega_c}(0)\right\}|0\rangle _\gamma \,,\\
\Pi_{\mu\nu}(p,q)&=i\int d^4x e^{ip \cdot x} \langle0|T\left\{J_\mu^{\Omega_c^*}(x)\bar{J}_\nu^{\Omega_c^*}(0)\right\}|0\rangle _\gamma \,,
\end{align}
where $T$ is the time ordered product; and the $J^{\Omega_c}(x)$ and  $J_\mu^{\Omega_c^*}(x)$ are the interpolating currents for the $\Omega_c$ states. The explicit form of the $J^{\Omega_c}(x)$ and $J_\mu^{\Omega_c^*}(x)$ are written as follows 
\begin{align}
 \label{curpcs1}
J^{\Omega_c}(x)&=\varepsilon^{abc}\varepsilon^{ade}\varepsilon^{bfg}\Big\{\big[  q^{d^T}_{1}(x) C\gamma_\mu q_2^e(x)\big] \big[q_3^{f^T}(x) 
 C\gamma^\mu c^g(x)\big] 
 C\bar{d}^{c^T}(x)\Big\},
\\
J_\mu^{\Omega_c^*}(x)&=\varepsilon^{abc}\varepsilon^{ade}\varepsilon^{bfg}\Big\{\big[  q^{d^T}_{1}(x) C\gamma_\mu q_2^e(x)\big] \big[q_3^{f^T}(x) 
 C\gamma_5 c^g(x)\big] 
 C\bar{d}^{c^T}(x)\Big\},
 \label{curpcs2}
\end{align}
where   
the $C$ is the charge conjugation operator and $a$, $b$... are color indices. The quark content of the $\Omega_c$ states is presented in Table \ref{quarkcon}. It should be noted here that the mass values obtained with these interpolating currents are compatible with the mass values of the $\Omega_c(3050)$, $\Omega_c(3066)$, $\Omega_c(3090)$, and $\Omega_c(3119)$ states~\cite{Wang:2021cku,Wang:2018alb}. 
\begin{table}[htp]
	\addtolength{\tabcolsep}{10pt}
		\begin{center}
		\caption{The quark content of the $\Omega_c$ states.}
	\label{quarkcon}
\begin{tabular}{l|c|c|ccc}
	   \hline\hline
  States&  $q_1$&$q_2$&$q_3$\\
\hline\hline
	   $\big[ss\big]\big[d c\big] \bar d$&  s & s& d\\
	   $\big[sd\big]\big[s c\big] \bar d$&  s & d& s\\
	   $\big[ds\big]\big[s c\big] \bar d$&  d & s& s\\
	   \hline\hline
\end{tabular}
\end{center}
\end{table}

After this brief introduction to the methodology, we begin the analysis of electromagnetic form factors in terms of hadronic parameters, following the above procedure. To evaluate the hadronic side, a complete set of hadronic states with the same quantum numbers is inserted into the correlation functions with the considered interpolating currents. Integration over x is then performed, yielding
 \begin{align}
 \label{edmn02}
 \Pi^{Had}(p,q)&=\frac{\langle0\mid  J^{\Omega_c}(x)\mid
{\Omega_c}(p,s)\rangle}{[p^{2}-m_{{\Omega_c}}^{2}]}
\langle {\Omega_c}(p,s)\mid
{\Omega_c}(p+q,s)\rangle_\gamma
\frac{\langle {\Omega_c}(p+q,s)\mid
\bar{J}^{\Omega_c}(0)\mid 0\rangle}{[(p+q)^{2}-m_{{\Omega_c}}^{2}]}+ \cdots , \\
\Pi^{Had}_{\mu\nu}(p,q)&=\frac{\langle0\mid  J_{\mu}^{\Omega_c^*}(x)\mid
{\Omega_c^*}(p,s)\rangle}{[p^{2}-m_{{\Omega_c^*}}^{2}]}
\langle {\Omega_c^*}(p,s)\mid
{\Omega_c^*}(p+q,s)\rangle_\gamma
\frac{\langle {\Omega_c^*}(p+q,s)\mid
\bar{J}_{\nu}^{\Omega_c^*}(0)\mid 0\rangle}{[(p+q)^{2}-m_{{\Omega_c^*}}^{2}]}+ \cdots ,\label{Pc103}
\end{align}
where $\cdots$ stands for the continuum and higher states. 
The Eqs.~(\ref{edmn02}) and (\ref{Pc103}) contain matrix elements containing hadronic parameters like residues ($\lambda$), form factors ($F_i$), spinors ($u_{(\mu)}$) and so on. The explicit expressions for these matrix elements are as follows
\begin{align}
\langle0\mid J^{\Omega_c}(x)\mid {\Omega_c}(p, s)\rangle=&\lambda_{\Omega_c} \gamma_5 \, u(p,s),\label{edmn04}\\
\langle {\Omega_c}(p+q, s)\mid\bar J^{\Omega_c}(0)\mid 0\rangle=&\lambda_{\Omega_c} \gamma_5 \, \bar u(p+q,s)\label{edmn004}
,\\
\langle {\Omega_c}(p, s)\mid {\Omega_c}(p+q, s)\rangle_\gamma &=\varepsilon^\mu\,\bar u(p, s)\bigg[\big[f_1(q^2)
+f_2(q^2)\big] \gamma_\mu +f_2(q^2)
\frac{(2p+q)_\mu}{2 m_{\Omega_c}}\bigg]\,u(p+q, s), \label{edmn005}
\\
\langle0\mid J_{\mu}^{\Omega_c^*}(x)\mid {\Omega_c^*}(p,s)\rangle&=\lambda_{{\Omega_c^*}}u_{\mu}(p,s),\\
\langle {\Omega_c^*}(p+q,s)\mid
\bar{J}_{\nu}^{\Omega_c^*}(0)\mid 0\rangle &= \lambda_{{\Omega_c^*}}\bar u_{\nu}(p+q,s), \\
\langle {\Omega_c^*}(p,s)\mid {\Omega_c^*}(p+q,s)\rangle_\gamma &=-e\bar
u_{\mu}(p,s)\bigg[F_{1}(q^2)g_{\mu\nu}\eslash 
-
\frac{1}{2m_{{\Omega_c^*}}} 
\Big[F_{2}(q^2)g_{\mu\nu} 
+F_{4}(q^2)\frac{q_{\mu}q_{\nu}}{(2m_{{\Omega_c^*}})^2}\Big]\eslash\qslash
\nonumber\\
&+
F_{3}(q^2)\frac{1}{(2m_{{\Omega_c^*}})^2}q_{\mu}q_{\nu}\eslash \bigg] 
u_{\nu}(p+q,s).
\label{matelpar}
\end{align}
%
%

With the help of the above formulas and the application of some mathematical simplifications, the following result is derived for the hadronic side of the correlation functions,
\begin{align}
\label{edmn0005}
\Pi^{Had}(p,q)=&\lambda^2_{\Omega_c}\gamma_5 \frac{\big(\pslash+m_{\Omega_c} \big)}{[p^{2}-m_{{\Omega_c}}^{2}]}\varepsilon^\mu \bigg[\big[f_1(q^2) %
+f_2(q^2)\big] \gamma_\mu
+f_2(q^2)\, \frac{(2p+q)_\mu}{2 m_{\Omega_c}}\bigg]  \gamma_5 
\frac{\big(\pslash+\qslash+m_{\Omega_c}\big)}{[(p+q)^{2}-m_{{\Omega_c}}^{2}]}, \\
  \Pi^{Had}_{\mu\nu}(p,q)&=-\frac{\lambda_{_{\Omega_c^*}}^{2}\,\big(\pslash+\qslash+m_{\Omega_c^*}\big)}{[(p+q)^{2}-m_{_{\Omega_c^*}}^{2}][p^{2}-m_{_{\Omega_c^*}}^{2}]}
 \bigg[g_{\mu\nu}
-\frac{1}{3}\gamma_{\mu}\gamma_{\nu}-\frac{2\,p_{1\mu}p_{1\nu}}
{3\,m^{2}_{\Omega_c^*}}+\frac{(p+q)_{\mu}\gamma_{\nu}-(p+q)_{\nu}\gamma_{\mu}}{3\,m_{\Omega_c^*}}\bigg]   \nonumber\\
& \times  \bigg\{F_{1}(q^2)g_{\mu\nu}\eslash-
\frac{1}{2m_{\Omega_c^*}}
\Big[F_{2}(q^2)g_{\mu\nu} +F_{4}(q^2) \frac{q_{\mu}q_{\nu}}{(2m_{\Omega_c^*})^2}\Big]\eslash\qslash+\frac{F_{3}(q^2)}{(2m_{\Omega_c^*})^2}
 q_{\mu}q_{\nu}\eslash\bigg\}
 \nonumber\\
 &\times
 \big(\pslash+m_{\Omega_c^*}\big)
 \bigg[g_{\mu\nu}-\frac{1}{3}\gamma_{\mu}\gamma_{\nu}-\frac{2\,p_{\mu}p_{\nu}}
{3\,m^{2}_{\Omega_c^*}}+\frac{p_{\mu}\gamma_{\nu}-p_{\nu}\gamma_{\mu}}{3\,m_{\Omega_c^*}}\bigg].
\label{final phenpart}
\end{align}

In order to express the hadronic representation in a more clear and straightforward manner, it is possible to present the above equations as follows:
\begin{align}
\label{edmn050}
\Pi^{Had}(p,q)&=\frac{\lambda^2_{\mathrm{\Omega_c}}}{[(p+q)^2-m^2_{\mathrm{\Omega_c}}][p^2-m^2_{\mathrm{\Omega_c}}]}
  \bigg[\Big(f_1(q^2)+f_2(q^2)\Big) \pslash\eslash\qslash + \cdots \bigg],\\
\Pi^{Had}_{\mu\nu}(p,q)&=\frac{\lambda_{_{{\mathrm{\Omega_c^*}}}}^{2}}{[(p+q)^{2}-m_{_{{\mathrm{\Omega_c^*}}}}^{2}][p^{2}-m_{_{{\mathrm{\Omega_c^*}}}}^{2}]} 
\bigg[ \varepsilon_\mu q_\nu \qslash\,F_{1}(q^2) 
+g_{\mu\nu}\pslash\eslash\qslash\,F_{2}(q^2)
+
\cdots 
\bigg]. \label{final phenpart1}
\end{align}

The magnetic dipole moments $(\mu)$ are determined from magnetic form factors at zero momentum transfer,
 \begin{eqnarray}\label{mqo2}
 \mu_{\Omega_c}&=&\frac{e}{2m_{\Omega_c}}F_{M}(0),~~~~~~\\
\mu_{\Omega_c^*}&=&\frac{e}{2m_{\Omega_c^*}}G_{M}(0),~~~~~~
\end{eqnarray}
where  $F_{M}(0)$ and $G_{M}(0)$ stand for magnetic form factor of the spin-$\frac{1}{2}$  and spin-$\frac{3}{2}$ $\Omega_c$ states, respectively.  Explicit forms of these terms are given as,
\begin{eqnarray}\label{mqo1}
F_{M}(0)&=&f_{1}(0)+f_{2}(0),\\
G_{M}(0)&=&F_{1}(0)+F_{2}(0).
\end{eqnarray}

Now that we have finished calculating the analysis regarding hadronic parameters, it is time to calculate in terms of QCD parameters. The magnetic dipole moments are analyzed in terms of QCD parameters using the interpolating current in the correlation function.  Applying Wick's theorem, we contract all the light and heavy quark fields to get the desired outcomes:
\begin{align}
\Pi^{\rm{QCD}}(p,q)=&-i\,\varepsilon^{abc}\varepsilon^{a^{\prime}b^{\prime}c^{\prime}}\varepsilon^{ade}
\varepsilon^{a^{\prime}d^{\prime}e^{\prime}}\varepsilon^{bfg}
\varepsilon^{b^{\prime}f^{\prime}g^{\prime}}\int d^4x e^{ip\cdot x} \langle 0| 
\bigg\{
\rm{Tr}\Big[\gamma^\mu  S_c^{gg^\prime}(x) \gamma^\nu  \widetilde S_{q_3}^{ff^\prime}(x)\Big]
\rm{Tr}\Big[\gamma_\mu  S_{q_2}^{ee^\prime}(x) \gamma_\nu \widetilde S_{q_1}^{dd^\prime}(x)\Big] \nonumber\\
&-
\rm{Tr}\Big[\gamma^\mu  S_c^{gg^\prime}(x) \gamma^\nu  \widetilde S_{q_3}^{ff^\prime}(x)\Big]
Tr\Big[\gamma_\mu  S_{q_2 q_1}^{ed^\prime}(x) \gamma_\nu \widetilde S_{q_1 q_2}^{de^\prime}(x)\Big]  
-  \rm{Tr}\Big[\gamma^\mu  S_c^{gg^\prime}(x) \gamma^\nu  \widetilde S_{q_2 q_3}^{ef^\prime}(x) 
\gamma_\mu  S_{q_1}^{dd^\prime}(x) \gamma_\nu \widetilde S_{q_3 q_2}^{fe^\prime}(x)\Big] \nonumber\\
&-
  \rm{Tr} \Big[\gamma^\mu  S_c^{gg^\prime}(x) \gamma^\nu  \widetilde S_{q_1 q_3}^{df^\prime}(x) 
\gamma_\mu \widetilde S_{q_2}^{ee^\prime}(x) \gamma_\nu \widetilde S_{q_3 q_1}^{fd^\prime}(x)\Big]
+
 \rm{Tr} \Big[\gamma^\mu  S_c^{gg^\prime}(x) \gamma^\nu  \widetilde S_{q_1 q_3}^{df^\prime}(x) 
\gamma_\mu  S_{q_2 q_1}^{ed^\prime}(x) \gamma_\nu \widetilde S_{q_3 q_2}^{fe^\prime}(x)\Big]\nonumber\\
&+
\rm{Tr} \Big[\gamma^\mu  S_c^{gg^\prime}(x) \gamma^\nu  \widetilde S_{q_2 q_3}^{ef^\prime}(x) 
\gamma_\mu  S_{q_1 q_2}^{de^\prime}(x) \gamma_\mu \widetilde S_{q_3 q_1}^{fd^\prime}(x)\Big]
\bigg \} \widetilde S_d^{c^{\prime}c}(-x)
|0 \rangle_\gamma,\label{QCD}
\end{align}

\begin{align}
\Pi^{\rm{QCD}}_{\mu\nu}(p,q)=&-i\,\varepsilon^{abc}\varepsilon^{a^{\prime}b^{\prime}c^{\prime}}\varepsilon^{ade}
\varepsilon^{a^{\prime}d^{\prime}e^{\prime}}\varepsilon^{bfg}
\varepsilon^{b^{\prime}f^{\prime}g^{\prime}}\int d^4x e^{ip\cdot x} \langle 0| 
\bigg\{
\rm{Tr}\Big[\gamma_5  S_c^{gg^\prime}(x) \gamma_5  \widetilde S_{q_3}^{ff^\prime}(x)\Big]
\rm{Tr}\Big[\gamma_\mu  S_{q_2}^{ee^\prime}(x) \gamma_\nu \widetilde S_{q_1}^{dd^\prime}(x)\Big] \nonumber\\
&-
\rm{Tr}\Big[\gamma_5  S_c^{gg^\prime}(x) \gamma_5  \widetilde S_{q_3}^{ff^\prime}(x)\Big]
Tr\Big[\gamma_\mu  S_{q_2 q_1}^{ed^\prime}(x) \gamma_\nu \widetilde S_{q_1 q_2}^{de^\prime}(x)\Big]  
-  \rm{Tr}\Big[\gamma_5  S_c^{gg^\prime}(x) \gamma_5  \widetilde S_{q_2 q_3}^{ef^\prime}(x) 
\gamma_\mu  S_{q_1}^{dd^\prime}(x) \gamma_\nu \widetilde S_{q_3 q_2}^{fe^\prime}(x)\Big] \nonumber\\
&-
  \rm{Tr} \Big[\gamma_5  S_c^{gg^\prime}(x) \gamma_5  \widetilde S_{q_1 q_3}^{df^\prime}(x) 
\gamma_\mu \widetilde S_{q_2}^{ee^\prime}(x) \gamma_\nu \widetilde S_{q_3 q_1}^{fd^\prime}(x)\Big]
+
 \rm{Tr} \Big[\gamma_5  S_c^{gg^\prime}(x) \gamma_5  \widetilde S_{q_1 q_3}^{df^\prime}(x) 
\gamma_\mu  S_{q_2 q_1}^{ed^\prime}(x) \gamma_\nu \widetilde S_{q_3 q_2}^{fe^\prime}(x)\Big]\nonumber\\
&+
\rm{Tr} \Big[\gamma_5  S_c^{gg^\prime}(x) \gamma_5  \widetilde S_{q_2 q_3}^{ef^\prime}(x) 
\gamma_\mu  S_{q_1 q_2}^{de^\prime}(x) \gamma_\mu \widetilde S_{q_3 q_1}^{fd^\prime}(x)\Big]
\bigg \} \widetilde S_d^{c^{\prime}c}(-x)
|0 \rangle_\gamma,\label{QCD1}
\end{align}
where   
$\widetilde{S}_{c(q)}^{ij}(x)=CS_{c(q)}^{ij\rm{T}}(x)C$. $S_{q_{i}q_{j}}$ exists when $q_i=q_j$ but it vanishes when $q_i\neq q_j$.  The $S_{c}(x)$ and $S_{q}(x)$ are the heavy and light-quark propagators whose expressions in the coordinate space are given as~\cite{Yang:1993bp, Belyaev:1985wza},
\begin{align}
\label{edmn13}
S_{q}(x)&= S_q^{free}(x) 
- \frac{\langle \bar qq \rangle }{12} \Big(1-i\frac{m_{q} \xslash}{4}   \Big)
- \frac{ \langle \bar qq \rangle }{192}
m_0^2 x^2  \Big(1 
  -i\frac{m_{q} \xslash}{6}   \Big)
+\frac {i g_s~G^{\mu \nu} (x)}{32 \pi^2 x^2} 
\Big[\rlap/{x} 
\sigma_{\mu \nu} +  \sigma_{\mu \nu} \rlap/{x}
 \Big],\\
%
S_{c}(x)&=S_c^{free}(x)
-\frac{m_{c}\,g_{s}\, G^{\mu \nu}(x)}{32\pi ^{2}} \bigg[ (\sigma _{\mu \nu }{\xslash}
+{\xslash}\sigma _{\mu \nu }) 
    \frac{K_{1}\big( m_{c}\sqrt{-x^{2}}\big) }{\sqrt{-x^{2}}}
 +2\sigma_{\mu \nu }K_{0}\big( m_{c}\sqrt{-x^{2}}\big)\bigg],
 \label{edmn14}
\end{align}%
with  
\begin{align}
 S_q^{free}(x)&=\frac{1}{2 \pi x^2}\Big(i \frac{\xslash}{x^2}- \frac{m_q}{2}\Big),\\
 S_c^{free}(x)&=\frac{m_{c}^{2}}{4 \pi^{2}} \Bigg[ \frac{K_{1}\big(m_{c}\sqrt{-x^{2}}\big) }{\sqrt{-x^{2}}}
+i\frac{{\xslash}~K_{2}\big( m_{c}\sqrt{-x^{2}}\big)}
{(\sqrt{-x^{2}})^{2}}\Bigg],
\end{align}
where $m_q$ and $\langle \bar qq \rangle$ stand for the light-quark mass and condensates, respectively, $G^{\mu\nu}$ being the gluon field-strength tensor, and $K_0$, $K_1$, and $K_2$ for the modified second type Bessel functions. 

The correlation function in Eqs.~(\ref{QCD}) and (\ref{QCD1})  would both receive perturbative and non-perturbative contributions. The interactions between photon and light or heavy quarks at short distances are referred to as perturbative contributions, whilst interactions between photon and quarks at large distances are referred to as non-perturbative contributions. To ensure the completeness and reliability of the magnetic dipole moment results obtained from the QCD light-cone sum rule method, it is necessary to include contributions from both regions. For the evaluation of the perturbative contributions we use the following formula:
\begin{align}
\label{free}
S^{free}(x) \rightarrow \int d^4z\, S^{free} (x-z)\,\rlap/{\!A}(z)\, S^{free} (z)\,,
\end{align}
where one of the propagators for either the light or heavy quark interacts with the photon at a short distance, and the surviving four propagators are assumed to be free. In the case of the non-perturbative contributions, the following equation is to be used
 \begin{align}
\label{edmn21}
S_{\alpha\beta}^{ab}(x) \rightarrow -\frac{1}{4} \big[\bar{q}^a(x) \Gamma_i q^b(0)\big]\big(\Gamma_i\big)_{\alpha\beta},
\end{align}
where it is assumed that the four surviving propagators are full, with one of the quarks interacting with the photon at a large distance. Here $\Gamma_i = \{\textbf{1}$, $\gamma_5$, $\gamma_\mu$, $i\gamma_5 \gamma_\mu$, $\sigma_{\mu\nu}/2\}$. Substituting Eq.~(\ref{edmn21}) into Eqs.~(\ref{QCD}) and (\ref{QCD1}) yields matrix elements $\langle \gamma(q)\vel \bar{q}(x) \Gamma_i G_{\alpha\beta}q(0) \ver 0\rangle$ and $\langle \gamma(q)\vel \bar{q}(x) \Gamma_i q(0) \ver 0\rangle$, which are defined in terms of the photon distribution amplitudes. These matrix elements, expressed by photon wave functions, play a crucial role in the evaluation of non-perturbative contributions (See Ref.~\cite{Ball:2002ps} for details on the distribution amplitudes (DAs) of the photon).   It should be noted that the photon DAs used in this study only take into account the contributions of the light quarks. However, in principle, photons can be emitted at a long distance from the c-quark.  The matrix elements of nonlocal operators are proportional to the quark condensates, the product of DAs, and some constants.  In the case of c-quark, the contribution of non-perturbative constants is negligible and can be neglected. The c-quark condensates are proportional to $1/m_c$. The condensates for the c-quark are to a large extent suppressed because of their large mass~\cite{Antonov:2012ud}.   In our calculations, we will therefore not use DAs with c-quark.  We have only taken into account the short-distance photon emission from the c-quark. By following the technical schemes mentioned above, we arrive at the QCD representation of the analysis of the magnetic dipole moments.  Details of the procedure used to obtain the expression of the perturbative and non-perturbative contributions can be found in Refs.~\cite{Ozdem:2022eds,Ozdem:2022vip} for those who are interested.

 After completing the procedures described above, we have finished analytical calculations of the magnetic dipole moments for the relevant states. For simplicity, analytical calculations for the spin-1/2 $\Omega_c$ states are presented as an example in the appendix.
The results for the magnetic dipole moments of $\Omega_c$ states are presented in the following section through numerical analysis.

%
%
%

\end{widetext}

\section{Numerical results} \label{secIII}

In this section, we numerically study the magnetic dipole moments of $\Omega_c$ states. The physical parameters to be calculated are analyzed using the following parameters: 
$m_s =93.4^{+8.6}_{-3.4}\,\mbox{MeV}$,
$m_c = 1.27\pm 0.02\,$GeV~\cite{ParticleDataGroup:2022pth}, $\langle \bar qq\rangle $=$(-0.24\pm 0.01)^3\,$GeV$^3$,  $\langle \bar ss\rangle $= $0.8 \langle \bar qq\rangle$~\cite{Ioffe:2005ym},   
$m_0^{2} = 0.8 \pm 0.1$~GeV$^2$, and $\langle g_s^2G^2\rangle = 0.48 \pm 0.14~ $GeV$^4$~\cite{Narison:2018nbv}.  In numerical analysis, we set $m_u$ =$m_d$ = 0 and $m^2_s = 0$, but consider terms proportional to $m_s$. 
The mass and residue values for these states have been calculated using the two-point sum rules of QCD, which have been used in our analysis~\cite{Wang:2021cku,Wang:2018alb}. The photon distribution amplitudes and the parameters utilized in these distribution amplitudes are borrowed from Ref.~\cite{Ball:2002ps}.

Beyond the parameters listed above, the QCD light-cone sum rules method includes two other parameters known as the Borel mass parameter $\rm{M^2}$ and the continuum threshold $\rm{s_0}$, and in the ideal scenario, these two parameters are expected to have a very small impact on the results. To do this, it is necessary to find the working regions of these parameters, where the variation of the results with respect to these parameters should be minimal. The $\rm{s_0}$ is the point at which the correlation function begins to include contributions from both excited states and the continuum.  To determine the working region of this parameter, the assumption  $\rm{s_0} = (M_H + 0.4^{+0.1}_{-0.1})~\rm{GeV }^2$ is usually adopted. The results are then analyzed for their dependence on slight variations of this parameter. We search for the ideal Borel parameters $\rm{M^2}$ according to the two criteria: pole contribution (PC) and OPE convergence (CVG). 
%
%
 These two conditions are both satisfied in the region
 $11.4~\mbox{GeV}^2 \leq \rm{s_0} \leq 12.8~\mbox{GeV}^2$ and  
  $3.0~\mbox{GeV}^2 \leq \rm{M^2} \leq 4.2~\mbox{GeV}^2$ for the $\Omega_c$ states; and $11.9~\mbox{GeV}^2 \leq \rm{s_0} \leq 13.3~\mbox{GeV}^2$ and  $3.2~\mbox{GeV}^2 \leq \rm{M^2} \leq 4.8~\mbox{GeV}^2$ for the $\Omega_c^*$ states. 
%
Our numerical evaluations show that considering these working intervals for the $\rm{M^2}$, the magnetic dipole moments of the $\Omega_c$ states PC vary within the interval $34\%\leq \rm{PC} \leq 61\%$ and $31\%\leq \rm{PC} \leq 57\%$ for $\Omega_c$ and $\Omega_c^*$ states, respectively. Analyzing the CVG, one sees that the contribution of the higher twist and higher dimensional terms in the OPE is $<3 \%$ of the total and the series shows good convergence for both $\Omega_c$ and $\Omega_c^*$ states. As an example, for fixed values of the continuum threshold $\rm{s_0}$, the variation of the extracted magnetic dipole moments of the  spin-$\frac{1}{2}^-$ $\Omega_c$ states with Borel mass $\rm{M^2}$ is shown in Fig.1. As it is seen from this figure, the magnetic dipole moments of these states exhibit a slight dependence on the parameter $\rm{M^2}$. While the magnetic dipole moments of the $\Omega_c$ states exhibit some dependence on $\rm{s_0}$, this is within the limits authorized by the approach and constitutes the main part of the uncertainties.

After setting all the relevant parameters, we are finally ready to estimate the numerical values of the corresponding magnetic dipole moments of $\Omega_c$ states. The predicted results for the magnetic dipole moments are shown in Table~\ref{sonuc}. The errors of the results, caused by errors in the input parameters and uncertainties in the estimation of the working intervals of the auxiliary parameters, are also shown in Table \ref{sonuc}.  
  %
    \begin{table}[htp]
	\addtolength{\tabcolsep}{10pt}
	\caption{ Results of the magnetic dipole moments of the $\Omega_c$ states.}
	\label{sonuc}
		\begin{center}
		\scalebox{0.9}{
\begin{tabular}{l|ccccccc}
                \hline\hline
              \\
 Parameter& $J^P$ & $ ssdc \bar d$ &  $ sdsc \bar d$& $dssc \bar d$ \\
     \\
                                        \hline\hline
                                      \\
 $\mu [\mu_N]$& $\frac{1}{2}^-$&$0.17^{+0.11}_{-0.08} $ & $0.62^{+0.14}_{-0.14} $ & $ 0.61^{+0.15}_{-0.13} $              
                        \\
    \\
                                        \hline\hline
                                      \\
 $\mu [\mu_N]$& $\frac{3}{2}^-$ &$-0.058^{+0.032}_{-0.031} $ & $-0.37^{+0.05}_{-0.07} $ & $-0.37^{+0.05}_{-0.07} $                                        
                       \\
                         \\
            \hline\hline
 \end{tabular}
}
\end{center}
\end{table}
The order of magnetic dipole moments can yield insights into their experimental measurability. The analysis of the obtained results indicates that the extracted magnetic dipole moments for both spin-1/2 and spin-3/2 $\Omega_c$ states can be experimentally measured. Our analysis shows that although hadrons have the same quark content, different diquark components can significantly affect the magnetic dipole moment results. This indicates that the magnetic dipole moment is a crucial parameter in determining the internal structure of hadrons. As previously stated, the interpolating currents used for both spin-1/2 and spin-3/2 $\Omega_c$ states predicted compatible with mass values. However, the magnetic dipole moments obtained for these states have different magnitudes and signs. This information may help us understand their quantum numbers and internal structure in experimental measurements. Comparing the magnetic dipole moment results obtained in this study with those extracted using different theoretical approaches can provide insight into the consistency of our predictions. The measurement of the magnetic dipole moments of unstable hadrons is a challenging experimental task. Nevertheless, using models approximating key features of QCD can give clues to the relevant dynamical mechanisms behind the observed structure. Together with spectroscopic/decay parameters and electromagnetic properties, it will also be interesting to characterize the branching ratios of the different decay modes and decay channels of $\Omega_c$ states.

Finally, this section briefly discusses the measurement of the magnetic dipole moment of the $\Omega_c$ states in the experiment.  Due to their short lifetimes, measuring the magnetic dipole moments of $\Omega_c$ states by spin precession experiments is a challenging task.  Instead, the magnetic dipole moment of these unstable states can only be measured indirectly through a three-step process. The particle is first produced, followed by the emission of a low-energy photon that acts as an external magnetic field. As the final step, the particle decays. There is a procedure for the indirect determination of the magnetic dipole moments of these states, which is based on the soft photon emission of hadrons proposed in Ref.~\cite{Zakharov:1968fb}. The basic concept underlying the procedure is that the photon carries information about the magnetic dipole and other higher multipole moments of the emitted particle. 
From this point of view, we can determine the magnetic dipole moment of the particles under consideration through measuring the cross-section or decay width of the radiative process. 
The magnetic dipole moment of states under investigation can affect the cross sections, whether total or differential. Determining the dipole moment of states under investigation requires comparing theoretical predictions with measured cross-sections.  The magnetic dipole moment of $\Delta(1232)$ resonance has been achieved by using the experimental data obtained in the $\gamma N \rightarrow  \Delta(1232) \rightarrow  \Delta(1232) \gamma \rightarrow \pi N \gamma $ process employing this procedure~\cite{Pascalutsa:2004je, Pascalutsa:2005vq, Pascalutsa:2007wb,   Kotulla:2002cg,Drechsel:2001qu,Drechsel:2000um,Chiang:2004pw,Machavariani:2005vn}.

\section{summary and outlook}\label{secIV}

Electromagnetic form factors, which describe the response of composite particles to electromagnetic surveys, provide an important tool for understanding the structure of bound states in QCD. The magnetic dipole moment is the leading-order response
of a bound system to a weak external magnetic field. It therefore offers an exquisite laboratory for studying the internal structure of composite particles, which is governed by the quark-gluon dynamics of QCD. Therefore, we systematically study the magnetic dipole moment of controversial states whose internal structure is not elucidated and we try to offer a different point of view to unravel the internal structure of these states.  Inspired by the $\Omega_c$ states observed by the LHCb Collaboration, we focus on the scenario of the diquark-diquark-antiquark pentaquark interpretation of the $\Omega_c$ states with both $J^P = \frac{1}{2}^-$ and $J^P = \frac{3}{2}^-$  quantum numbers, and study the electromagnetic properties of these states with the QCD light-cone sum rules by employing the distribution amplitudes of the photon. From the obtained numerical results, we conclude that the magnetic dipole moments of the $\Omega_c$ states can reflect their inner structures, which can be used to distinguish their spin-parity quantum numbers. 
The order of the magnetic dipole moments obtained for these states implies that they are accessible by experiments.
Measuring the magnetic dipole moments of the $\Omega_c$ states in future experimental facilities can be very helpful for understanding the internal organization and identifying the quantum numbers of these states.  We hope that in the future we will have more experimental data on a variety of physical quantities of these states.

\begin{widetext}

\begin{figure}[htp]
\centering
 \subfloat[]{\includegraphics[width=0.4\textwidth]{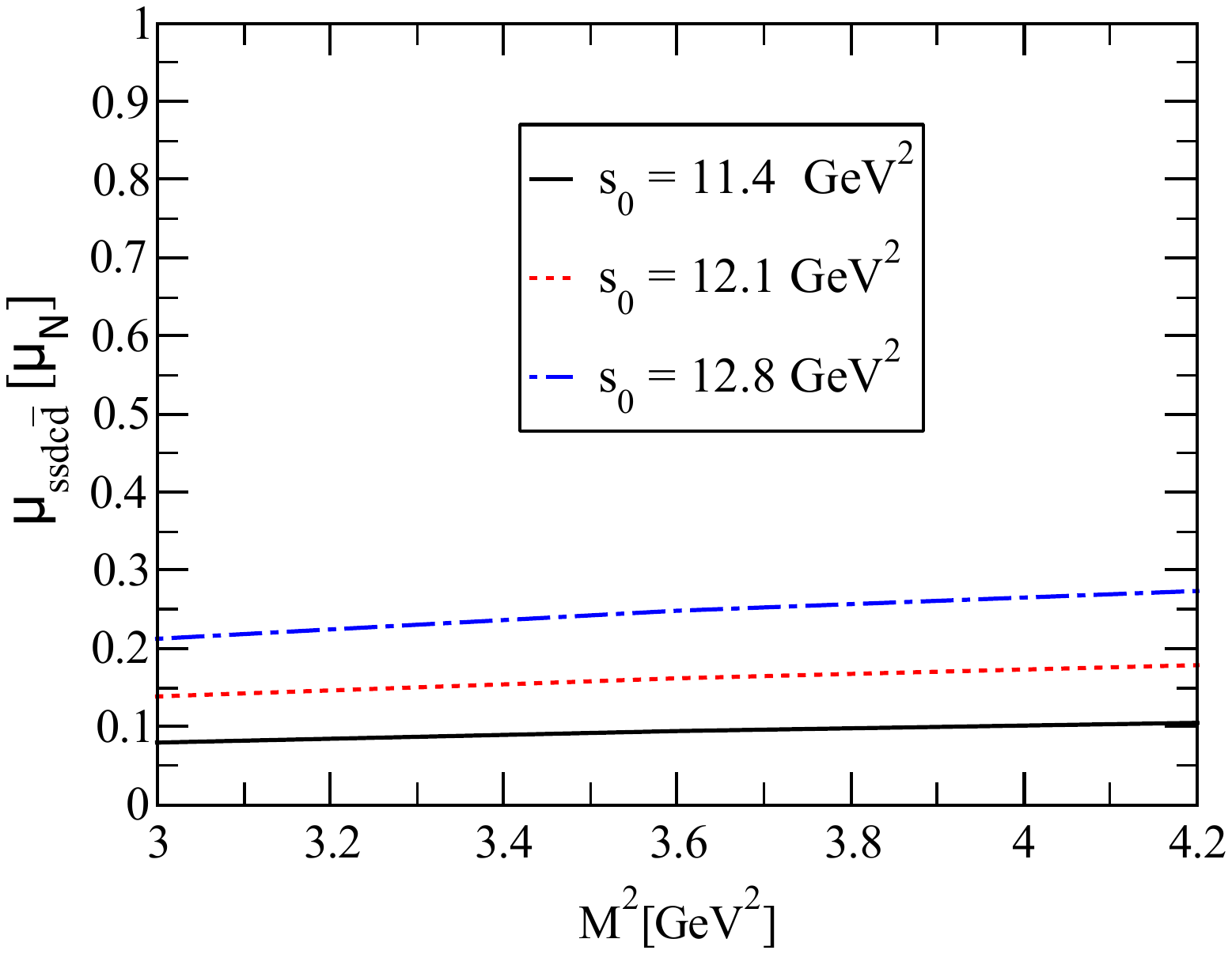}}~~~~~~~~
 \subfloat[]{\includegraphics[width=0.4\textwidth]{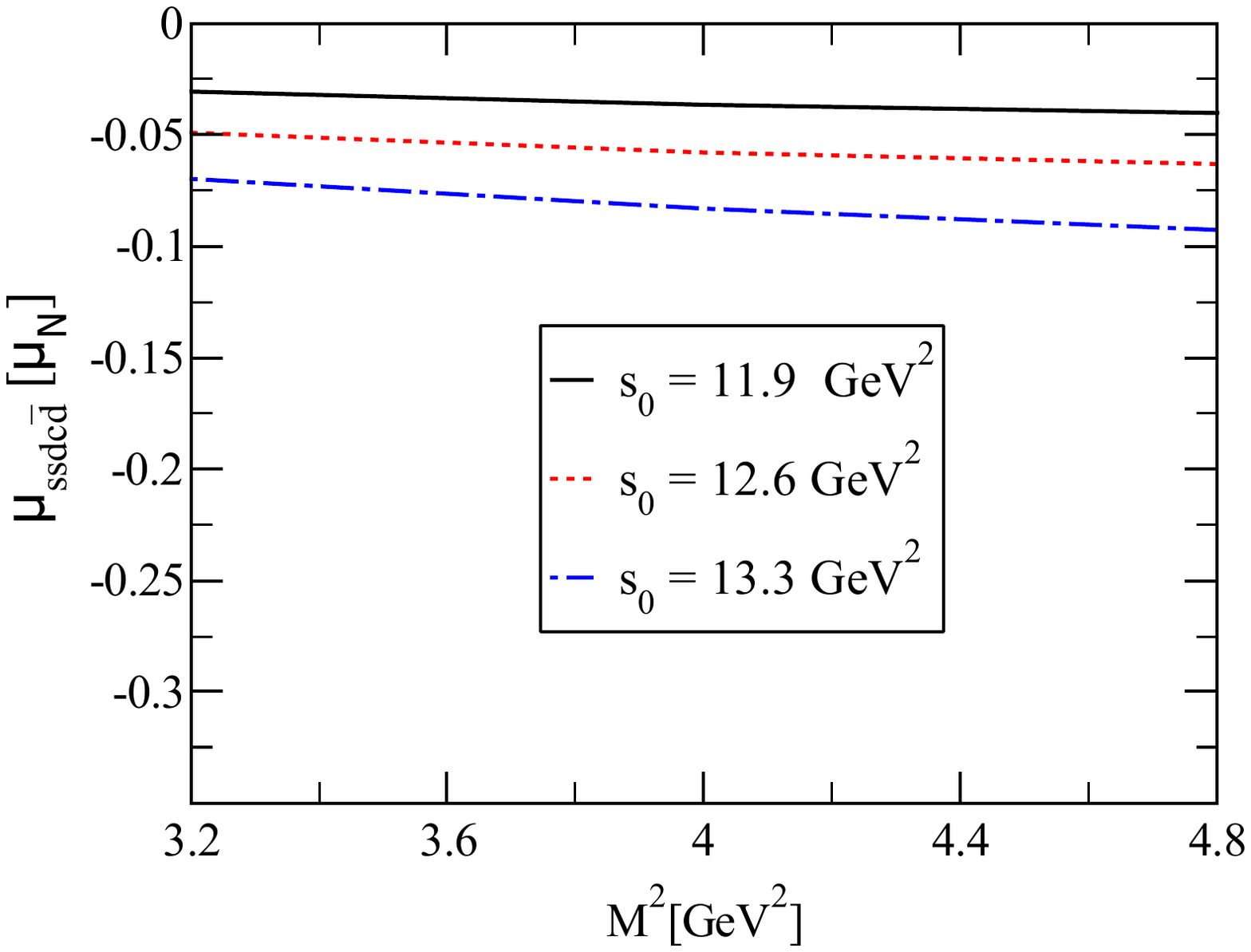}}
\\
 \subfloat[]{\includegraphics[width=0.4\textwidth]{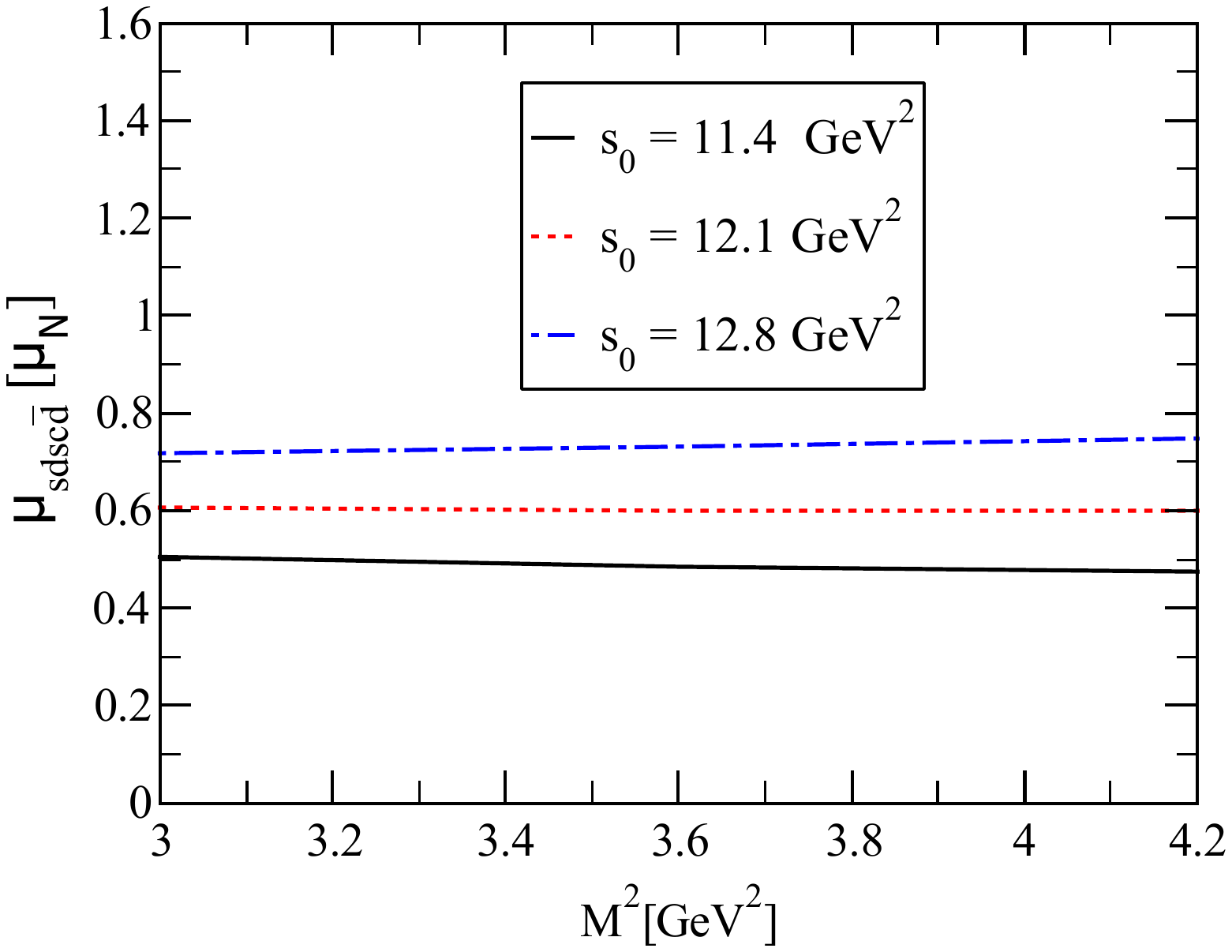}}~~~~~~~~
 \subfloat[]{\includegraphics[width=0.4\textwidth]{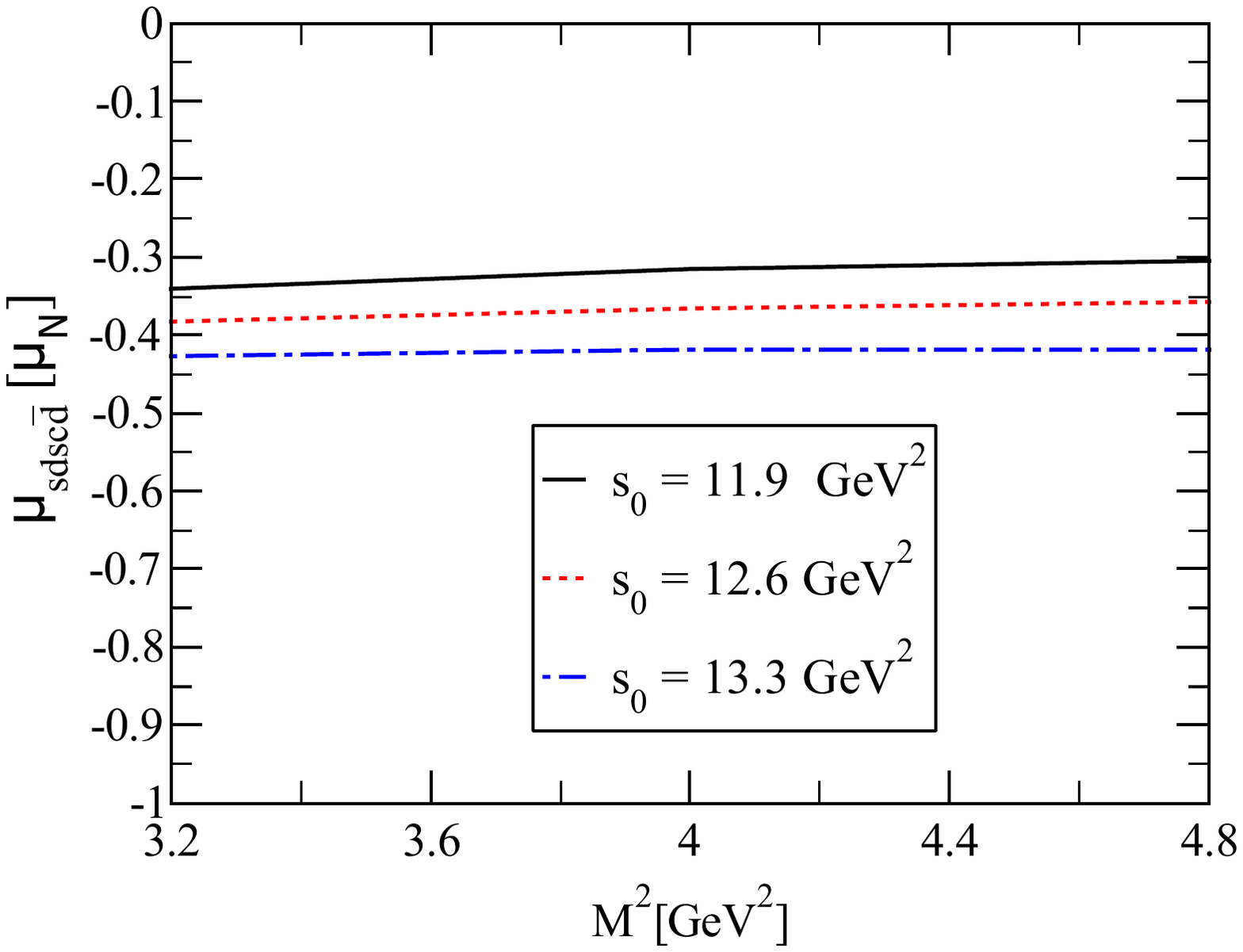}}
 \\
 \subfloat[]{\includegraphics[width=0.4\textwidth]{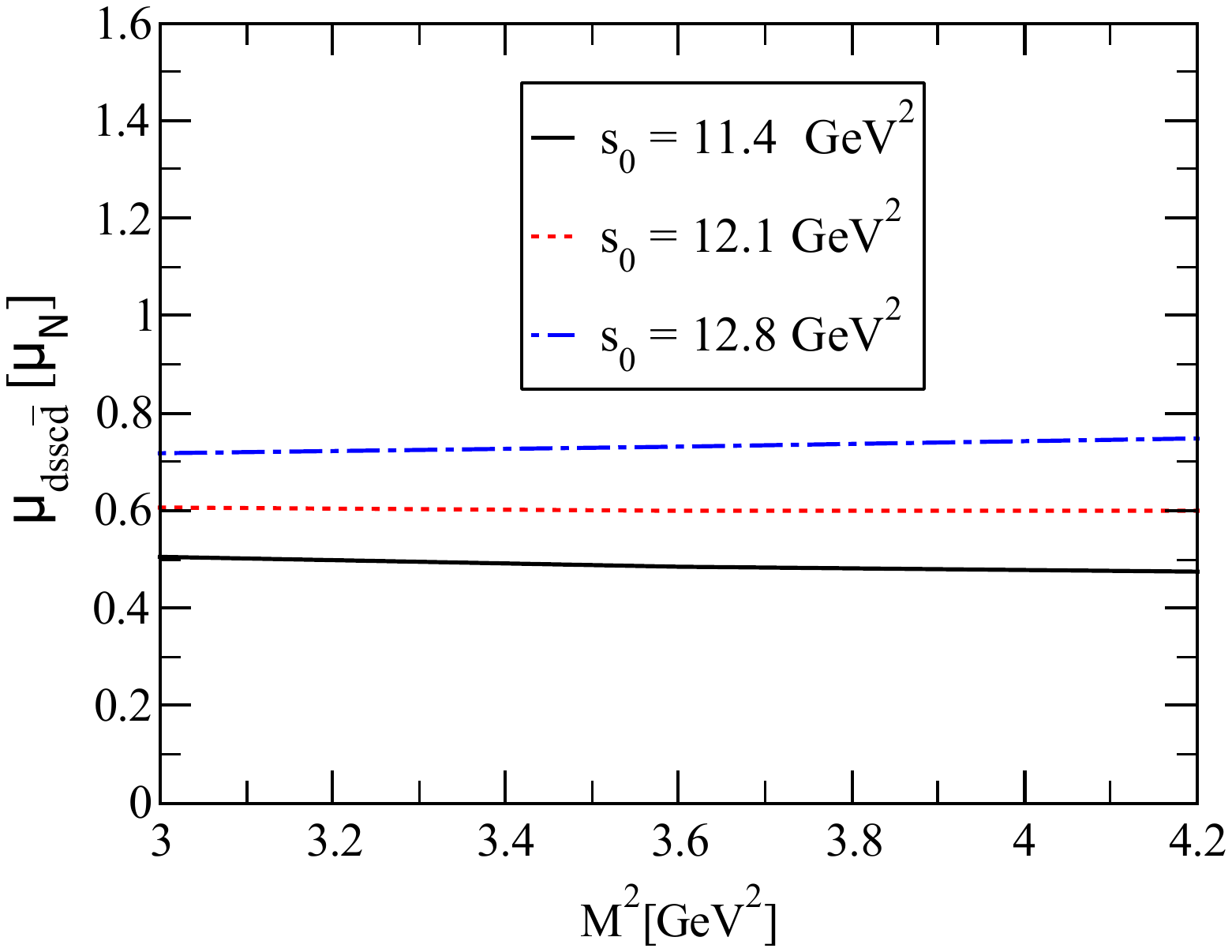}}~~~~~~~~
 \subfloat[]{\includegraphics[width=0.4\textwidth]{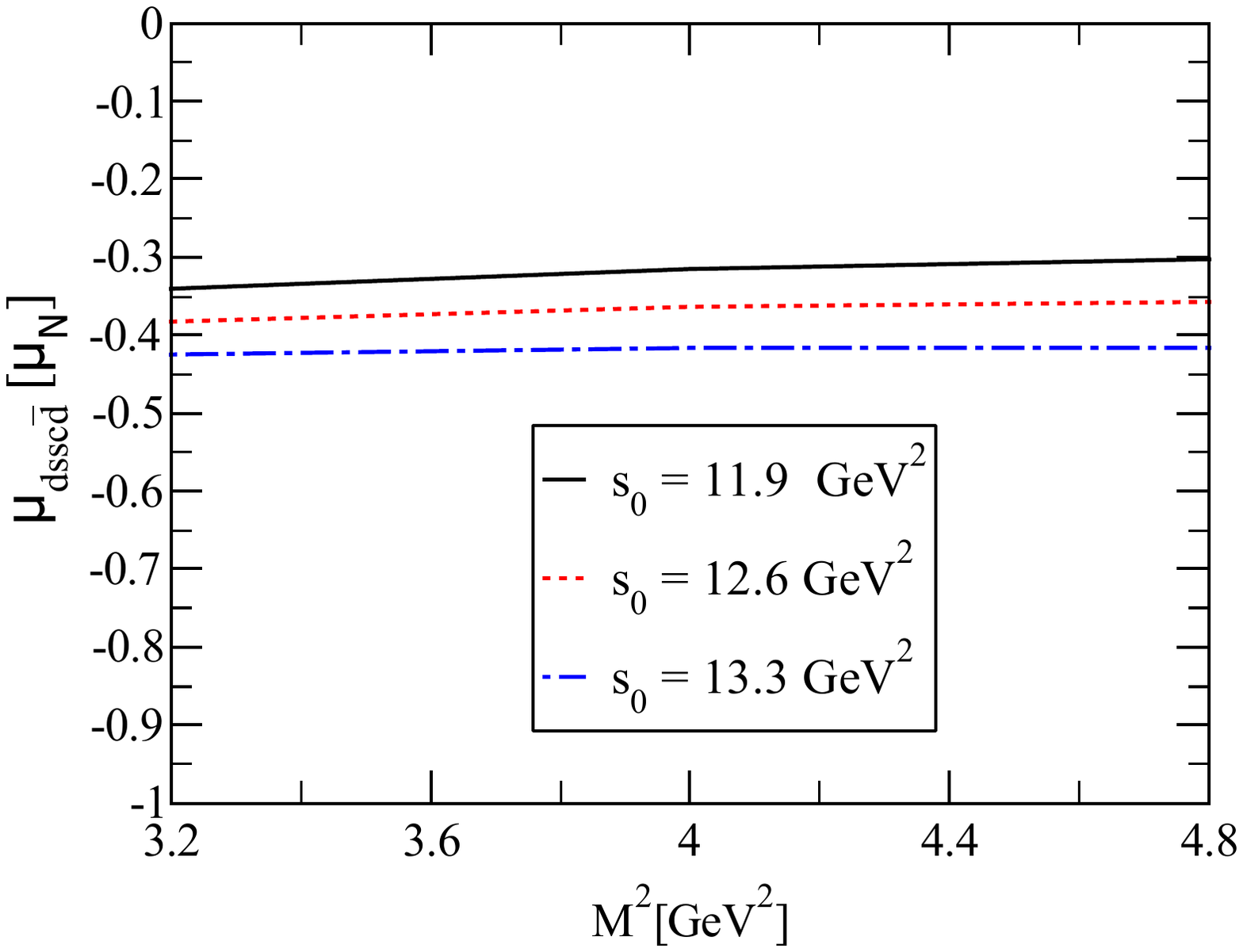}}
 \caption{ Dependencies of the magnetic dipole moments of  $\Omega_c$ states on $\rm{M^2}$ at three values of $\rm{s_0}$; (a), (c), and (e) for spin-$\frac{1}{2}$ states, and; (b), (d), and (f) for spin-$\frac{3}{2}$ states. }
 \label{Msqfig}
  \end{figure}
  
  \end{widetext}
\begin{widetext}
%
\section*{Appendix: Analytical results of the magnetic dipole moments of the spin-1/2 $\Omega_c$ states}
\begin{align}
 \mu_{\Omega_c}&=\frac{e^{\frac{m^2_{\Omega_c}}{\rm{M^2}}}}{\,\lambda^2_{\Omega_c}}\Bigg\{ 
 -\frac{e_d}{2^{13} \times 3 \times  5 m_c^2 \pi^8}\Bigg[ 
 2 m_c^{14} I[-7, 5] + 5 m_{q_3} m_c^{13} I[-6, 4] + 
 m_c^{12} \big(-10 m_{q_1} m_{q_2} I[-6, 4] + 8 I[-6, 5]\big) \nonumber\\
 &+ 
 20 m_{q_3} m_c^{11} \big(4 m_{q_1} m_{q_2} I[-5, 3] - I[-5, 4]\big) + 
 6 m_c^{10} \big(5 m_{q_1} m_{q_2} I[-5, 4] + 2 I[-5, 5]\big) + 
 30 m_{q_3} m_c^9 \big(8 m_{q_1} m_{q_2} I[-4, 3]\nonumber\\
 & + I[-4, 4]\big) + 
 m_c^8 \big(-30 m_{q_1} m_{q_2} I[-4, 4] + 8 I[-4, 5]\big) + 
 20 m_{q_3} m_c^7 \big(12 m_{q_1} m_{q_2} I[-3, 3] - I[-3, 4]\big) \nonumber\\
 &+ 
 2 m_c^6 \big(5 m_{q_1} m_{q_2} I[-3, 4] + I[-3, 5]\big) + 
 5 m_{q_3} m_c^5 \big(16 m_{q_1} m_{q_2} I[-2, 3] + I[-2, 4]\big) + 
 640 m_{q_1} m_{q_2} m_{q_3} m_c I[0, 3] \nonumber\\
 &+ 32 I[0, 5]\Bigg]\nonumber\\
 &-\frac {P_ 1 P_ 3} {2^{18}  3^4 m_c^2 \pi^6} \Bigg[
 -54 e_ {q_ 1} m_c  \big (m_c^6 I[-3, 1] + 2 m_c^4 I[-2, 1] + 
   m_c^2 I[-1, 1] + 4 I[0, 1]\big) I_ 3[\mathcal {\tilde S}]
   -3 e_ {q_ 1} m_c  \big(81 m_c^6 I[-3, 1] \nonumber\\
   &+ 159 m_c^4 I[-2, 1] + 
   184 m_ {q_ 2} m_c^3 (m_c^2 I[-3, 1] + I[-2, 1]) - 
   156 m_ {q_ {23}} m_c^3 (m_c^2 I[-3, 1] + I[-2, 1]) + 
   78 m_c^2 I[-1, 1] \nonumber\\
   &+ 312 I[0, 1]\big) I_ 3[\mathcal S]
   +8 f_ {3\gamma}\pi^2  \big (26 e_ {q_ 3} m_ {q_ 1} (m_c^4 I[-2, 0] - 
      I[0, 0]) + 
   27 e_ {q_ {23}} m_c^3 (-I[-1, 0] + 
       m_c^2 I[2, 0])\big) I_ 2[\mathcal V]\nonumber\\
       &
       +24 e_ {q_ 1}  \Big (4 m_ {q_ 2}\big (3 m_c^6 I[-3, 1] + 
       2 m_c^4 I[-2, 1] - 2 m_ {q_ 3} m_c^3 I[-1, 0] + I[0, 1] + 
       2 m_ {q_ 3} m_c^5 I[2, 0]\big) - 
    3 \big (m_c^7 I[-3, 1] \nonumber\\
       &- m_c^3 I[-1, 1] + 
        2 m_ {q_ {23}} (m_c^6 I[-3, 1] + 2 m_c^4 I[-2, 1] - 
           I[0, 1]) - 4 m_c I[0, 1]\big)\Big)\mathbb {A}[u_ 0]
           +48 \chi e_ {q_ 1} m_c \Big (4 m_c^8 I[-4, 2] \nonumber\\
           &+ 
    m_c^6 (8 m_ {q_ 2} m_ {q_ 3} I[-3, 1] - 9 I[-3, 2]) + 
    4 m_ {q_ 2} m_c^5 I[-3, 2] - 4 m_ {q_ 2} m_c^3 I[-2, 2] + 
    6 m_ {q_ {23}} m_c^3 (2 m_c^4 I[-4, 2] \nonumber\\
    &- 3 m_c^2 I[-3, 2] + 
       I[-2, 2]) + 
    2 m_c^4 (8 m_ {q_ 2} m_ {q_ 3} I[-2, 1] + 3 I[-2, 2]) + 
    m_c^2 (8 m_ {q_ 2} m_ {q_ 3} I[-1, 1] - I[-1, 2]) \nonumber\\
    &+ 
    32 m_ {q_ 2} m_ {q_ 3} I[0, 1]\Big) \varphi_ {\gamma}[u_ 0]
    +128 f_ {3\gamma}\pi^2 \Big (2 e_ {q_ 2} m_ {q_ 1} (m_c^4 I[-2, 0] - 
       I[0, 0]) + 4 e_ {q_ 3} m_ {q_ 2} (m_c^4 I[-2, 0] - I[0, 0]) \nonumber\\
       &+ 
    3 e_ {q_ {23}} \big (2 m_c^3 I[-1, 0] + m_c I[0, 0] + 
        m_ {q_ {23}} (-2 m_c^4 I[-2, 0] + 2 I[0, 0]) - 
        m_c^5 (I[-2, 0] + 2 I[2, 0])\big)\Big) \psi^a[u_ 0]
 \Bigg]\nonumber\\
 &-\frac {P_ 1 P_ 4} {2^{18}  3^4 m_c^2 \pi^6} \Bigg[
  -54 e_ {q_ 2} m_c  \big (m_c^6 I[-3, 1] + 2 m_c^4 I[-2, 1] + 
   m_c^2 I[-1, 1] + 4 I[0, 1]\big) I_ 3[\mathcal {\tilde S}]
   -3 e_ {q_ 2} m_c  \big(81 m_c^6 I[-3, 1] \nonumber\\
   &+ 159 m_c^4 I[-2, 1] + 
   184 m_ {q_ 2} m_c^3 (m_c^2 I[-3, 1] + I[-2, 1]) - 
   156 m_ {q_ {13}} m_c^3 (m_c^2 I[-3, 1] + I[-2, 1]) + 
   78 m_c^2 I[-1, 1] \nonumber\\
   &+ 312 I[0, 1]\big) I_ 3[\mathcal S]
   +8 f_ {3\gamma}\pi^2  \big (26 e_ {q_ 3} m_ {q_ 2} (m_c^4 I[-2, 0] - 
      I[0, 0]) + 
   27 e_ {q_ {13}} m_c^3 (-I[-1, 0] + 
       m_c^2 I[2, 0])\big) I_ 2[\mathcal V]\nonumber\\
       &
       +24 e_ {q_ 2}  \Big (4 m_ {q_ 2}\big (3 m_c^6 I[-3, 1] + 
       2 m_c^4 I[-2, 1] - 2 m_ {q_ 3} m_c^3 I[-1, 0] + I[0, 1] + 
       2 m_ {q_ 3} m_c^5 I[2, 0]\big) - 
    3 \big (m_c^7 I[-3, 1] \nonumber\\
       &- m_c^3 I[-1, 1] + 
        2 m_ {q_ {13}} (m_c^6 I[-3, 1] + 2 m_c^4 I[-2, 1] - 
           I[0, 1]) - 4 m_c I[0, 1]\big)\Big)\mathbb {A}[u_ 0]
           +48 \chi e_ {q_ 2} m_c \Big (4 m_c^8 I[-4, 2] \nonumber\\
       &+ 
    m_c^6 (8 m_ {q_ 2} m_ {q_ 3} I[-3, 1] - 9 I[-3, 2]) + 
    4 m_ {q_ 2} m_c^5 I[-3, 2] - 4 m_ {q_ 2} m_c^3 I[-2, 2] + 
    6 m_ {q_ {13}} m_c^3 (2 m_c^4 I[-4, 2] \nonumber\\
       &- 3 m_c^2 I[-3, 2] + 
       I[-2, 2]) + 
    2 m_c^4 (8 m_ {q_ 2} m_ {q_ 3} I[-2, 1] + 3 I[-2, 2]) + 
    m_c^2 (8 m_ {q_ 2} m_ {q_ 3} I[-1, 1] - I[-1, 2]) \nonumber\\
       &+ 
    32 m_ {q_ 2} m_ {q_ 3} I[0, 1]\Big) \varphi_ {\gamma}[u_ 0]
    +128 f_ {3\gamma}\pi^2 \Big (2 e_ {q_ 2} m_ {q_ 2} (m_c^4 I[-2, 0] - 
       I[0, 0]) + 4 e_ {q_ 3} m_ {q_ 2} (m_c^4 I[-2, 0] - I[0, 0]) \nonumber\\
       &+ 
    3 e_ {q_ {13}} \big (2 m_c^3 I[-1, 0] + m_c I[0, 0] + 
        m_ {q_ {13}} (-2 m_c^4 I[-2, 0] + 2 I[0, 0]) - 
        m_c^5 (I[-2, 0] + 2 I[2, 0])\big)\Big) \psi^a[u_ 0]
 \Bigg]\nonumber\\
  &+\frac {P_ 1 P_ 5} {2^{15}  3^4 m_c^2 \pi^6} \Bigg[-24 e_ {q_ 3} m_c^3 I_ 3[\mathcal {\tilde S}] I[-1, 
   1] - (e_ {q_ 1} + 2 e_ {q_ {12}} + 
    e_ {q_ 2}) f_ {3\gamma} \pi^2 I_ 2[\mathcal V] \Big (41 m_c^3 \
I[-1, 0] + 26 m_ {q_ 3} (m_c^4 I[-2, 0])\Big)\nonumber\\
&
+3 e_ {q_ 3} m_c \Big (-8  \big (3 m_c^6 I[-3, 1] + 4 m_c^4 I[-2, 1] +
        m_c^2 I[-1, 1] + 4 I[0, 1]\big)\mathbb A[
     u_ 0] + \big (75 m_c^6 I[-3, 1] + 127 m_c^4 I[-2, 1] \nonumber\\
     &+ 
      52 m_c^2 I[-1, 1] + 208 I[0, 1]\big) I_ 3[\mathcal S] \Big)
      -32 (e_ {q_ 1} + 2 e_ {q_ {12}} + 
   e_ {q_ 2}) f_ {3\gamma} \pi^2  \Big (2 m_c^3 I[-1, 0] + 
    m_ {q_ 3} (m_c^4 I[-2, 0] - I[0, 0]) \nonumber\\
     & - m_c I[0, 0] + 
    m_c^5 (I[-2, 0] - 2 I[2, 0])\Big) \psi^a[u_ 0]
    +16 \chi e_ {q_ 3} m_c^3 \Big (2 m_c^6 I[-4, 2] - 3 m_c^4 I[-3, 2] + 
    I[-1, 2]\Big)\varphi_{\gamma}[u_ 0]\Bigg]\nonumber
            \end{align}
 \begin{align}
    &-\frac {P_ 1 P_ 6} {2^{17}  3^4 m_c^2 \pi^6} \Bigg[
    -54 e_ {q_ {12}} m_c \big (m_c^6 I[-3, 1] + 2 m_c^4 I[-2, 1] + 
     m_c^2 I[-1, 1] + 4 I[0, 1]\big) I_ 3[\mathcal {\tilde S}] - 
  3 e_ {q_ {12}} m_c  \Big (-78 m_ {q_ {13}} m_c^5 I[-3, 1]\nonumber\\
     & + 
     81 m_c^6 I[-3, 1] - 78 m_ {q_ {23}} m_c^3 I[-2, 1] + 
     159 m_c^4 I[-2, 1] + 
     184 m_ {q_ {12}} m_c^3 (m_c^2 I[-3, 1] + I[-2, 1]) - 
     78 m_ {q_ {13}} m_c^3 (m_c^2 I[-3, 1] \nonumber\\
     &+ I[-2, 1]) + 
     78 m_c^2 I[-1, 1] + 312 I[0, 1]\Big) I_ 3[\mathcal S] + 
  24 e_ {q_ {12}}  \Bigg (4 m_ {q_ {12}} (3 m_c^6 I[-3, 1] + 
         2 m_c^4 I[-2, 1] - 2 m_ {q_ {3}} m_c^3 I[-1, 0] + I[0, 1] 
         \nonumber\\
     &+ 
         2 m_ {q_ {3}} m_c^5 I[2, 0]) - 
      3 \Big (m_ {q_ {23}} m_c^6 I[-3, 1] + m_c^7 I[-3, 1] + 
          2 m_ {q_ {23}} m_c^4 I[-2, 1] - m_c^3 I[-1, 1] - 
          m_ {q_ {23}} I[0, 1] - 4 m_c I[0, 1] \nonumber\\
     &+ 
          m_ {q_ {13}}\big (m_c^6 I[-3, 1] + 2 m_c^4 I[-2, 1] - 
              4 m_ {q_ {23}} m_c^3 I[-1, 0] - I[0, 1] + 
              4 m_ {q_ {23}} m_c^5 I[2, 0]\big)\Big)\Bigg)\mathbb A[
    u_ 0] \nonumber\\
     &+ 4 \Bigg (9 e_ {q_ {23}} f_ {3\gamma} m_c^3 \pi^2  (-I[-1, 
           0] + m_c^2 I[2, 0]) I_ 2[\mathcal A] + 
      f_ {3\gamma} \pi^2 \Big (52 eq3 m_ {q_ {12}} (m_c^4 I[-2, 0] - 
            I[0, 0]) + 
         27 (e_ {q_ {13}} + 
            e_ {q_ {23}}) m_c^3 \nonumber\\
     & \times (-I[-1, 0] + 
             m_c^2 I[2, 0])\Big) I_ 2[\mathcal V] + 
      12 \chi e_ {q_ {12}} m_c \Big (4 m_c^8 I[-4, 2] + 
          8 m_ {q_ {12}} m_ {q_ {3}} m_c^6 I[-3, 1] + 
          4 m_ {q_ {12}} m_c^5 I[-3, 2] - 9 m_c^6 I[-3, 2] \nonumber\\
     &+ 
          16 m_ {q_ {12}} m_ {q_ {3}} m_c^4 I[-2, 1] - 
          4 m_ {q_ {12}} m_c^3 I[-2, 2] + 6 m_c^4 I[-2, 2] + 
          3 m_ {q_ {23}} m_c^3 (2 m_c^4 I[-4, 2] - 3 m_c^2 I[-3, 2] + 
             I[-2, 2]) \nonumber\\
     &+ 8 m_ {q_ {12}} m_ {q_ {3}} m_c^2 I[-1, 1] - 
          m_c^2 I[-1, 2] + 32 m_ {q_ {12}} m_ {q_ {3}} I[0, 1] + 
          3 m_ {q_ {13}}\big (2 m_c^7 I[-4, 2] - 
              4 m_ {q_ {23}} m_c^6 I[-3, 1] - 3 m_c^5 I[-3, 2] \nonumber\\
     &- 
              8 m_ {q_ {23}} m_c^4 I[-2, 1] + m_c^3 I[-2, 2] - 
              4 m_ {q_ {23}} m_c^2 I[-1, 1] - 
              16 m_ {q_ {23}} I[0, 1]\big)\Big) \varphi_{\gamma}[
        u_ 0] + 16 f_{3\gamma} \pi]^2 \Big (4 e_ {q_ {12}} m_ {q_{12}} (m_c^4 I[-2, 0] \nonumber\\
     &- I[0, 0]) + 
          8 e_ {q_ 3} m_ {q_ {12}} (m_c^4 I[-2, 0] - I[0, 0]) + 
          3 e_ {q_ {13}} \big (2 m_c^3 I[-1, 0] + m_c I[0, 0] + 
              m_ {q_ {23}} (-2 m_c^4 I[-2, 0] + 2 I[0, 0]) \nonumber\\
     &- 
              m_c^5 (I[-2, 0] + 2 I[2, 0])\big)\Big) \psi^a[u_ 0]\Bigg)\Bigg]\nonumber\\
           &   +\frac {P_ 1 P_ 7} {2^{16}  3^3 m_c^2 \pi^6} \Bigg[
              3 e_ {q_ {13}} m_c  \Big (13 m_ {q_ {2}} m_c^5 I[-3, 1] - 
    m_ {q_ {23}} m_c^5 I[-3, 1] + 27 m_c^6 I[-3, 1] + 
    13 m_ {q_ {2}} m_c^3 I[-2, 1] - m_ {q_ {23}} m_c^3 I[-2, 1]\nonumber\\
     & + 
    41 m_c^4 I[-2, 1] + 
    13 m_ {q_ {12}} m_c^3 (m_c^2 I[-3, 1] + I[-2, 1]) - 
    m_ {q_ {13}} m_c^3 (m_c^2 I[-3, 1] + I[-2, 1]) + 
    14 m_c^2 I[-1, 1]  \nonumber\\
     & + 56 I[0, 1]\Big)I_ 3[\mathcal S] - 
 12 e_ {q_ {13}}  \Big (-m_ {q_ {2}} m_c^6 I[-3, 1] + 
    3 m_ {q_ {23}} m_c^6 I[-3, 1] - 2 m_c^7 I[-3, 1] - 
    2 m_ {q_ {2}} m_c^4 I[-2, 1] + 2 m_ {q_ {23}} m_c^4 I[-2, 1]\nonumber\\
     & - 
    4 m_c^5 I[-2, 1] - 2 m_c^3 I[-1, 1] + m_ {q_ {2}} I[0, 1] + 
    m_ {q_ {23}} I[0, 1] - 8 m_c I[0, 1] + 
    m_ {q_ {12}} (-m_c^6 I[-3, 1] - 2 m_c^4 I[-2, 1] + I[0, 1]) \nonumber\\
     &+ 
    m_ {q_ {13}} (3 m_c^6 I[-3, 1] + 2 m_c^4 I[-2, 1] + 
        I[0, 1])\Big) \mathbb A[u_0] - 
 2 \Bigg (f_ {3\gamma} \pi^2  \Big (16 (e_ {q_ {12}} + 
          e_ {q_ {2}}) m_c^3 (-I[-1, 0] + m_c^2 I[2, 0]) \nonumber\\
     &+ 
       e_ {q_ {13}} (m_ {q_ {13}} m_c^4 I[-2, 0] + 5 m_c^3 I[-1, 0] - 
          m_ {q_ {13}} I[0, 0] - 5 m_c^5 I[2, 0]) + 
       e_ {q_ {23}} (m_ {q_ {13}} m_c^4 I[-2, 0] + 5 m_c^3 I[-1, 0] - 
           m_ {q_ {13}} I[0, 0] \nonumber\\
     &- 
           5 m_c^5 I[2, 0])\Big) I_ 2[\mathcal V] + 
    4 \Big (e_ {q_ {23}} f_ {3\gamma} m_c^3 \pi^2  (I[-1, 0] - 
           m_c^2 I[2, 
             0]) I_ 2[\mathcal A] + \chi e_ {q_ {13}} m_c^3 \Big (-2 \
m_c^6 I[-4, 2] + 3 m_ {q_ {13}} m_c^3 I[-3, 2] \nonumber\\
     &+ 
            3 m_ {q_ {23}} m_c^3 I[-3, 2] + 6 m_c^4 I[-3, 2] - 
            3 m_ {q_ {13}} m_c I[-2, 2] - 
            3 m_ {q_ {23}} m_c I[-2, 2] - 6 m_c^2 I[-2, 2] + 
            3 m_ {q_ {12}} m_c (2 m_c^4 I[-4, 2] \nonumber\\
     &- 3 m_c^2 I[-3, 2] + 
               I[-2, 2]) + 
            3 m_ {q_ {2}} m_c (2 m_c^4 I[-4, 2] - 3 m_c^2 I[-3, 2] + 
               I[-2, 2]) + 2 I[-1, 2]\Big) \varphi_{\gamma}[u_ 0] + 
        4 f_ {3\gamma} \pi^2 \Big (e_ {q_ {12}}  \nonumber\\
     & \times \big (-2 m_ {q_ {13}}
m_c^4 I[-2, 0] + 2 m_ {q_ {23}} m_c^4 I[-2, 0] + m_c^5 I[-2, 0] - 
               2 m_c^3 I[-1, 0] + 2 m_ {q_ {13}} I[0, 0] - 
               2 m_ {q_ {23}} I[0, 0] - m_c I[0, 0] \nonumber\\
     & + 
               2 m_c^5 I[2, 0]\big)+ 
            e_ {q_ {13}} \big (-2 m_ {q_ {2}} m_c^4 I[-2, 0] - 
               m_c^5 I[-2, 0] - 2 m_c^3 I[-1, 0] + 
               2 m_ {q_ {2}} I[0, 0] + m_c I[0, 0] + 
               2 m_c^5 I[2, 0]\big)  \nonumber\\
     & + 
            e_ {q_ {2}} m_c \big (-2 m_c^2 I[-1, 0] - I[0, 0] + 
                m_c^4 (I[-2, 0] + 2 I[2, 0])\big)\Big) \psi^a[u_ 0]\Big)\Bigg)
              \Bigg]\nonumber\\
              &+\frac {P_ 1 P_ 8} {2^{17}  3^3 m_c^2 \pi^6} \Bigg[ 3 e_ {q_ {23}} m_c  \Big (27 m_c^6 I[-3, 1] + 41 m_c^4 I[-2, 1] + 
    26 m_ {q_ {1}} m_c^3 (m_c^2 I[-3, 1] + I[-2, 1]) - 
    2 m_ {q_ {23}} m_c^3 (m_c^2 I[-3, 1] \nonumber\\
     &+ I[-2, 1]) + 
    14 m_c^2 I[-1, 1] + 56 I[0, 1]\Big) I_ 3[\mathcal S] - 
 24 e_ {q_ {23}}  \Big (3 m_ {q_ {13}} m_c^6 I[-3, 1] + 
     3 m_ {q_ {23}} m_c^6 I[-3, 1] - 2 m_c^7 I[-3, 1] \nonumber\\
     &+ 
     2 m_ {q_ {13}} m_c^4 I[-2, 1] + 2 m_ {q_ {23}} m_c^4 I[-2, 1] - 
     4 m_c^5 I[-2, 1] - 2 m_c^3 I[-1, 1] + m_ {q_ {13}} I[0, 1] + 
     m_ {q_ {23}} I[0, 1] - 8 m_c I[0, 1] \nonumber\\
     &+ 
     m_ {q_ {1}} (-m_c^6 I[-3, 1] - 2 m_c^4 I[-2, 1] + I[0, 1]) + 
     m_ {q_ {12}} (-m_c^6 I[-3, 1] - 2 m_c^4 I[-2, 1] + 
         I[0, 1])\Big)\mathbb A[u_ 0]\nonumber
 \end{align}
 \begin{align}
      & - 
 4 \Bigg (f_ {3\gamma} \pi^2  \Big (16 (e_ {q_ {1}} + 
          e_ {q_ {12}}) m_c^3 (-I[-1, 0] + m_c^2 I[2, 0]) + 
       e_ {q_ {13}} (m_ {q_ {23}} m_c^4 I[-2, 0] + 5 m_c^3 I[-1, 0] - 
          m_ {q_ {23}} I[0, 0] \nonumber\\
     &- 5 m_c^5 I[2, 0]) + 
       e_ {q_ {23}} (m_ {q_ {23}} m_c^4 I[-2, 0] + 5 m_c^3 I[-1, 0] - 
           m_ {q_ {23}} I[0, 0] - 5 m_c^5 I[2, 0])\Big) + 
    4 \Big (\chi e_ {q_ {23}} m_c^3 \big (-2 m_c^6 I[-4, 2] \nonumber\\
     &+ 
            3 m_ {q_ {13}} m_c^3 I[-3, 2] + 
            3 m_ {q_ {23}} m_c^3 I[-3, 2] + 6 m_c^4 I[-3, 2] - 
            3 m_ {q_ {13}} m_c I[-2, 2] - 
            3 m_ {q_ {23}} m_c I[-2, 2] - 6 m_c^2 I[-2, 2] \nonumber\\
     &+ 
            3 m_ {q_ {1}} m_c (2 m_c^4 I[-4, 2] - 3 m_c^2 I[-3, 2] + 
               I[-2, 2]) + 
            3 m_ {q_ {12}} m_c (2 m_c^4 I[-4, 2] - 3 m_c^2 I[-3, 2] + 
               I[-2, 2])\nonumber\\
     & + 2 I[-1, 2]\big) \varphi_ {\gamma}[u_ 0] + 
        4 f_ {3\gamma} \pi^2 \big (2 e_ {q_ {12}} m_ {q_ {13}} m_c^4 \
I[-2, 0] - 2 e_ {q_ {12}} m_ {q_ {23}} m_c^4 I[-2, 0] + 
            e_ {q_ {1}} m_c^5 I[-2, 0] + 
            e_ {q_ {12}} m_c^5 I[-2, 0]\nonumber\\
     & - 
            2 e_ {q_ {1}} m_c^3 I[-1, 0] - 
            2 e_ {q_ {12}} m_c^3 I[-1, 0] - 
            2 e_ {q_ {12}} m_ {q_ {13}} I[0, 0] + 
            2 e_ {q_ {12}} m_ {q_ {23}} I[0, 0] - 
            e_ {q_ {1}} m_c I[0, 0] - e_ {q_ {12}} m_c I[0, 0] \nonumber\\
     &+ 
            2 e_ {q_ {1}} m_c^5 I[2, 0] + 
            2 e_ {q_ {12}} m_c^5 I[2, 0] + 
            e_ {q_ {23}} (-2 m_ {q_ {1}} m_c^4 I[-2, 0] - 
               m_c^5 I[-2, 0] - 2 m_c^3 I[-1, 0] + 
               2 m_ {q_ {1}} I[0, 0] + m_c I[0, 0] \nonumber\\
     &+ 
               2 m_c^5 I[2, 0]) + 
            e_ {q_ {13}} (-2 m_ {q_ {12}} m_c^4 I[-2, 0] - 
                m_c^5 I[-2, 0] - 2 m_c^3 I[-1, 0] + 
                2 m_ {q_ {12}} I[0, 0] + m_c I[0, 0] \nonumber\\
     &+ 
                2 m_c^5 I[2, 0])\big) \psi^a[u_ 0]\Big)\Bigg)
 \Bigg]\nonumber\\
  &-\frac { P_ 1 f_ {3\gamma}} {2^{19} 3^3 m_c \pi^6}\Bigg[  
 6 e_ {q_ {23}} (3 m_ {q_ {12}} - 
    4 m_ {q_ {13}})  \big (m_c^6 I[-3, 1] + 2 m_c^4 I[-2, 1] + 
   m_c^2 I[-1, 1] + 4 I[0, 1]\big) I_ 2[\mathcal A]\nonumber\\
   &+\Bigg (m_c^7 (12 e_ {q_ {13}} + 3 e_ {q_ 2} + 12 e_ {q_ {23}} + 
      208 e_ {q_ 3})  I[-4, 2] + 
   2 m_c^6 (3 e_ {q_ {13}} (9 m_ {q_ {12}} - 5 m_ {q_ {13}} + 
         9 m_ {q_ 2} - 5 m_ {q_ {23}}) + 
      3 e_ {q_ {23}} (9 m_ {q_ 1} \nonumber\\
       & + 9 m_ {q_ {12}}- 
         5 (m_ {q_ {13}} + m_ {q_ {23}})) + 
      e_ {q_ 2} (48 m_ {q_ {13}} - 41 m_ {q_ 3}))  I[-3, 1] - 
   m_c^5 (24 e_ {q_ {13}} + 55 e_ {q_ 2} + 24 e_ {q_ {23}} + 
      416 e_ {q_ 3})  I[-3, 2] \nonumber\\
   &+ 
   4 m_c^4 (3 e_ {q_ {13}} (9 m_ {q_ {12}} - 5 m_ {q_ {13}} + 
         9 m_ {q_ 2} - 5 m_ {q_ {23}}) + 
      3 e_ {q_ {23}} (9 m_ {q_ 1} + 9 m_ {q_ {12}} - 
         5 (m_ {q_ {13}} + m_ {q_ {23}})) + 
      e_ {q_ 2} (48 m_ {q_ {13}} - 41 m_ {q_ 3}))\nonumber\\
   & \times   I[-2, 1] + 
   4 m_c^3 (3 e_ {q_ {13}} + 13 e_ {q_ 2} + 3 e_ {q_ {23}} + 
      52 e_ {q_ 3})  I[-2, 2] + 208 e_ {q_ 3} m_c^3 I[-2, 2] + 
   2 m_c^2 (3 e_ {q_ {13}} (9 m_ {q_ {12}} - 5 m_ {q_ {13}} + 
         9 m_ {q_ 2} \nonumber\\
     &- 5 m_ {q_ {23}}) + 
      3 e_ {q_ {23}} (9 m_ {q_ 1} + 9 m_ {q_ {12}} - 
         5 (m_ {q_ {13}} + m_ {q_ {23}})) + 
      e_ {q_ 2} (48 m_ {q_ {13}} - 41 m_ {q_ 3}))  I[-1, 1] + 
   8 (3 e_ {q_ {13}} (9 m_ {q_ {12}} - 5 m_ {q_ {13}} + 9 m_ {q_ 2} \nonumber\\
   &- 
         5 m_ {q_ {23}}) + 
      3 e_ {q_ {23}} (9 m_ {q_ 1} + 9 m_ {q_ {12}} - 
         5 (m_ {q_ {13}} + m_ {q_ {23}})) + 
      e_ {q_ 2} (48 m_ {q_ {13}} - 41 m_ {q_ 3})) I[0, 1] + 
   2 e_ {q_ {12}} \Big (3 m_c^7 I[-4, 2] + 
      2 m_c^6 (24 m_ {q_ {13}}\nonumber\\
   & + 24 m_ {q_ {23}} - 
         41 m_ {q_ 3})  I[-3, 1] - 55 m_c^5 I[-3, 2] + 
      4 m_c^4 (24 m_ {q_ {13}} + 24 m_ {q_ {23}} - 
         41 m_ {q_ 3})  I[-2, 1] + 52 m_c^3 I[-2, 2] + 
      2 m_c^2 (24 m_ {q_ {13}}\nonumber\\
   & + 24 m_ {q_ {23}} - 
         41 m_ {q_ 3})  I[-1, 1] + 
      8 (24 m_ {q_ {13}} + 24 m_ {q_ {23}} - 41 m_ {q_ 3}) I[0, 
         1]\Big) + 
   e_ {q_ 1} \Big (3 m_c^7 I[-4, 2] + 
       2 m_c^6 (48 m_ {q_ {23}} - 41 m_ {q_ 3})  I[-3, 1] \nonumber\\
   &- 
       55 m_c^5 I[-3, 2] + 
       4 m_c^4 (48 m_ {q_ {23}} - 41 m_ {q_ 3})  I[-2, 1] + 
       52 m_c^3 I[-2, 2] + 
       2 m_c^2 (48 m_ {q_ {23}} - 41 m_ {q_ 3})  I[-1, 1] + 
       8 (48 m_ {q_ {23}} \nonumber\\
   &- 41 m_ {q_ 3}) I[0, 
          1]\Big)\Bigg) I_2[\mathcal V] 
   \Bigg]\nonumber\\
 &+\frac{f_{3\gamma}}{2^{16}3^2 m_c^2 \pi^6}\Bigg[\Bigg(m_c^{12}\big (12 e_ {q_ {13}} + e_ {q_ 2} + 12 e_ {q_ {23}} + 
    8 e_ {q_ 3}\big)  I[-6, 4] - 
 4 m_c^{10} \Big (6 e_ {q_ 3} m_ {q_ 1} m_ {q_ 2} + 
    9 e_ {q_ {13}} m_ {q_ {13}} m_ {q_ 2} + 
    9 e_ {q_ {13}} m_ {q_ {12}} m_ {q_ {23}}\nonumber\\
   & - 
    3 e_ {q_ {13}} m_ {q_ {12}} m_c + 
    3 e_ {q_ {13}} m_ {q_ {13}} m_c - 3 e_ {q_ 2} m_ {q_ {13}} m_c - 
    3 e_ {q_ {13}} m_ {q_ 2} m_c + 3 e_ {q_ {13}} m_ {q_ {23}} m_c + 
    2 e_ {q_ 2} m_ {q_ 3} m_c + 
    3 e_ {q_ {23}} \big (3 m_ {q_ {12}} m_ {q_ {13}} \nonumber\\
   &+ 
        3 m_ {q_ 1} m_ {q_ {23}} - m_ {q_ 1} m_c - m_ {q_ {12}} m_c + 
        m_ {q_ {13}} m_c + m_ {q_ {23}} m_c\big)\Big) I[-5, 3] - 
 3 m_c^10\big (12 e_ {q_ {13}} + e_ {q_ 2} + 12 e_ {q_ {23}} + 
    8 e_ {q_ 3}\big)  I[-5, 4]\nonumber\\
   &
-12 m_c^8 \Big (4 e_ {q_ 3} m_ {q_ 1} m_ {q_ 2} + 
    6 e_ {q_ {13}} m_ {q_ {13}} m_ {q_ 2} + 
    6 e_ {q_ {13}} m_ {q_ {12}} m_ {q_ {23}} - 
    3 e_ {q_ {13}} m_ {q_ {12}} m_c + 
    3 e_ {q_ {13}} m_ {q_ {13}} m_c - 3 e_ {q_ 2} m_ {q_ {13}} m_c - 
    3 e_ {q_ {13}} m_ {q_ 2} m_c \nonumber\\
   &+ 3 e_ {q_ {13}} m_ {q_ {23}} m_c + 
    2 e_ {q_ 2} m_ {q_ 3} m_c + 
    3 e_ {q_ {23}} \big (2 m_ {q_ {12}} m_ {q_ {13}} + 
        2 m_ {q_ 1} m_ {q_ {23}} - m_ {q_ 1} m_c - m_ {q_ {12}} m_c + 
        m_ {q_ {13}} m_c + m_ {q_ {23}} m_c\big)\Big) I[-4, 3] \nonumber\\
   &- 
 12 m_c^6 \Big (2 e_ {q_ 3} m_ {q_ 1} m_ {q_ 2} + 
    3 e_ {q_ {13}} m_ {q_ {13}} m_ {q_ 2} + 
    3 e_ {q_ {13}} m_ {q_ {12}} m_ {q_ {23}} - 
    3 e_ {q_ {13}} m_ {q_ {12}} m_c + 
    3 e_ {q_ {13}} m_ {q_ {13}} m_c - 3 e_ {q_ 2} m_ {q_ {13}} m_c - 
    3 e_ {q_ {13}} m_ {q_ 2} m_c \nonumber\\
   &+ 3 e_ {q_ {13}} m_ {q_ {23}} m_c + 
    2 e_ {q_ 2} m_ {q_ 3} m_c + 
    3 e_ {q_ {23}} \big (m_ {q_ {12}} (m_ {q_ {13}} - m_c) + 
        m_ {q_ 1} (m_ {q_ {23}} - m_c) + (m_ {q_ {13}} + 
           m_ {q_ {23}}) m_c\big)\Big) I[-3, 3]\nonumber\\
    & + 
 4 m_c^5 \big (3 e_ {q_ {23}} (m_ {q_ 1} + m_ {q_ {12}} - 
       m_ {q_ {13}} - m_ {q_ {23}}) + 
    3 e_ {q_ {13}} (m_ {q_ {12}} - m_ {q_ {13}} + m_ {q_ 2} - 
       m_ {q_ {23}}) + 
    e_ {q_ 2} (3 m_ {q_ {13}} - 2 m_ {q_ 3})\big)  I[-2, 3]\nonumber\\
   & - 
 16 \Big (6 e_ {q_ 3} m_ {q_ 1} m_ {q_ 2} + 
    9 e_ {q_ {13}} m_ {q_ {13}} m_ {q_ 2} + 
    9 e_ {q_ {13}} m_ {q_ {12}} m_ {q_ {23}} - 
    6 e_ {q_ {13}} m_ {q_ {12}} m_c + 
    6 e_ {q_ {13}} m_ {q_ {13}} m_c - 6 e_ {q_ 2} m_ {q_ {13}} m_c - 
    6 e_ {q_ {13}} m_ {q_ 2} m_c \nonumber\\
   &+ 6 e_ {q_ {13}} m_ {q_ {23}} m_c + 
    4 e_ {q_ 2} m_ {q_ 3} m_c + 
    e_ {q_ {23}} \big (9 m_ {q_ {12}} m_ {q_ {13}} + 
        9 m_ {q_ 1} m_ {q_ {23}} - 6 m_ {q_ 1} m_c - 
        6 m_ {q_ {12}} m_c + 6 m_ {q_ {13}} m_c + 
        6 m_ {q_ {23}} m_c\big)\Big) I[0, 3]\nonumber\\
   &
+e_ {q_ 1} m_c \Big (m_c^{11} I[-6, 4] + 
    4 (3 m_ {q_ {23}} - 2 m_ {q_ 3}) m_c^{10} I[-5, 3] - 
    3 m_c^9 I[-5, 4] + 
    12 (3 m_ {q_ {23}} - 2 m_ {q_ 3}) m_c^8 I[-4, 3] + 
    3 m_c^7 I[-4, 4] \nonumber
    \end{align}
    \begin{align}
   &+ 
    12 (3 m_ {q_ {23}} - 2 m_ {q_ 3}) m_c^6 I[-3, 3] - 
    m_c^5 I[-3, 4] + 
    4 (3 m_ {q_ {23}} - 2 m_ {q_ 3}) m_c^4 I[-2, 3] + 
    32 (3 m_ {q_ {23}} - 2 m_ {q_ 3}) I[0, 3]\Big) \nonumber\\
   &+ 
 2 e_ {q_ {12}} \Big (m_c^12 I[-6, 4] + 
    2 m_c^11 (3 m_ {q_ {13}} + 3 m_ {q_ {23}} - 4 m_ {q_ 3})  I[-5, 
      3] - 3 m_c^10 (6 m_ {q_ {13}} m_ {q_ {23}} I[-5, 3] + 
       I[-5, 4]) + 
    6 m_c^9 (3 m_ {q_ {13}} \nonumber\\
   &+ 3 m_ {q_ {23}} - 4 m_ {q_ 3})  I[-4, 
      3] + m_c^8 (-36 m_ {q_ {13}} m_ {q_ {23}} I[-4, 3] + 
       3 I[-4, 4]) + 
    6 m_c^7 (3 m_ {q_ {13}} + 3 m_ {q_ {23}} - 4 m_ {q_ 3})  I[-3, 
      3] \nonumber\\
   &- m_c^6 (18 m_ {q_ {13}} m_ {q_ {23}} I[-3, 3] + I[-3, 4]) + 
    2 m_c^5 (3 m_ {q_ {13}} + 3 m_ {q_ {23}} - 4 m_ {q_ 3})  I[-2, 
      3] - 72 m_ {q_ {13}} m_ {q_ {23}} I[0, 3] + 
    16 m_c (3 m_ {q_ {13}} \nonumber\\
   &+ 3 m_ {q_ {23}} - 4 m_ {q_ 3})  I[0, 
       3]\Big)
 \Bigg)I_2[\mathcal V]
 \Bigg]
 \Bigg\},
 \label{analitiksonuc}
\end{align}
where $P_1 = \langle g_s^2 G^2 \rangle $ is gluon condensate;   $P_3 = \langle \bar q_1 q_1 \rangle $, $P_4 = \langle \bar q_2 q_2 \rangle $, $P_5 = \langle \bar q_3 q_3 \rangle $, $P_6 = \langle \bar q_{12} q_{12} \rangle $, $P_7 = \langle \bar q_{13} q_{13} \rangle $ and $P_8 = \langle \bar q_{23} q_{23} \rangle $ are corresponding quark condensates.
 Note that only terms that contribute significantly to the numerical values of the magnetic dipole moments are included in the above expressions. For the sake of simplicity, higher dimensional contributions are not shown, although they are considered in the numerical calculations. For completeness, the values $e_{q_i}, e_{q_{ij}}$, $m_{q_i}, m_{q_{ij}}$, and $P_i$ related to the expressions of the magnetic moments in Eq.~(\ref{analitiksonuc}) are given in Table \ref{eqimqi}.
The $I[n,m]$,~$I_2[\mathcal{F}]$~and~$I_3[\mathcal{F}]$ functions
are presented as:
\begin{align}
 I[n,m]&= \int_{\mathcal M^2}^{\mathrm{s_0}} ds \int_{\mathcal M^2}^s dl~ e^{-s/\mathrm{M^2}}\,l^n~(s-l)^m ,\nonumber\\
 I_2[\mathcal{F}]&=\int D_{\alpha_i} \int_0^1 dv~ \mathcal{F}(\alpha_{\bar q},\alpha_q,\alpha_g)
 \delta(\alpha_{\bar q}+ v \alpha_g-u_0),\nonumber\\
 I_3[\mathcal{F}]&= \int_0^1 du \, \mathcal{F}(u), 
 \end{align}
 where $\mathcal M =m_c+2 m_s $, and $\mathcal{F}$ represents the corresponding photon DAs.
%
 \end{widetext}

    \begin{table}[htp]
	\addtolength{\tabcolsep}{10pt}
	\caption{ The values $e_{q_i}, e_{q_{ij}}$, $m_{q_i}, m_{q_{ij}}$, and $P_i$ related to the expressions of the magnetic moments in Eq.~(\ref{analitiksonuc}).}
	\label{eqimqi}
		\begin{center}
		\scalebox{0.85}{
\begin{tabular}{l|ccccccc}
                \hline\hline
 Parameters  & $ ssdc \bar d$ &  $ sdsc \bar d$& $dssc \bar d$ \\
                                        \hline\hline
 $e_{q_{1}}$& $e_s$ & $e_s$ & $ e_d $\\
  $e_{q_{2}}$&$e_s$ & $e_d$ & $ e_s $\\
   $e_{q_{3}}$&$e_d$ & $e_s$ & $ e_s $\\
    $e_{q_{12}}$& $e_s$ & $-$ & $ - $\\
     $e_{q_{13}}$ &$-$ & $e_s$ & $ - $\\
     $e_{q_{23}}$ &$-$ & $-$ & $ e_s $\\
                                        \hline\hline
$m_{q_{1}}$ &$m_s$ & $m_s$ & $ m_d $\\
  $m_{q_{2}}$ &$m_s$ & $m_d$ & $ m_s $\\
   $m_{q_{3}}$&$m_d$ & $m_s$ & $ m_s $\\
    $m_{q_{12}}$&$m_s$ & $-$ & $ - $\\
     $m_{q_{13}}$&$-$ & $m_s$ & $ - $\\
     $m_{q_{23}}$&$-$ & $-$ & $ m_s $\\
                \hline\hline
    $P_{3}$& $\langle \bar ss \rangle$ & $\langle \bar ss \rangle$ & $ \langle \bar qq \rangle $\\
      $P_{4}$ &$\langle \bar ss \rangle$ & $\langle \bar qq \rangle$ & $ \langle \bar ss \rangle $\\
       $P_{5}$& $\langle \bar qq \rangle$ & $\langle \bar ss \rangle$ & $ \langle \bar ss \rangle $\\
        $P_{6}$& $\langle \bar ss \rangle$ & $-$ & $ - $\\
         $P_{7}$& $-$ &$-$ & $\langle \bar ss \rangle$ \\
          $P_{8}$&$-$ & $-$ & $ \langle \bar ss \rangle $\\
     \hline\hline
 \end{tabular}
}
\end{center}
\end{table}

\bibliographystyle{elsarticle-num}
\bibliographystyle{apsrev4-1}
\bibliographystyle{apsrev4-2}
\bibliography{Newomega_c.bib}

\end{document}